\documentclass[journal]{IEEEtran}
\usepackage{amsmath}
\usepackage{amssymb}
\usepackage{multirow,tabularx}
\usepackage{booktabs}
\usepackage{graphicx}
\usepackage{epsfig}
\usepackage{times}
\usepackage{epsfig}
\usepackage{graphicx}
\usepackage{amsmath}
\usepackage{verbatim}
\usepackage{amssymb}
\usepackage{multirow}
\usepackage{booktabs}

\begin{document}
\title{An Efficient and Integrated Algorithm for Video Enhancement in Challenging Lighting Conditions}

\author{Xuan~Dong,
        Jiangtao~(Gene)~Wen,~\IEEEmembership{Senior~Member,~IEEE,}
        Weixin~Li,~Yi~(Amy)~Pang,~Guan~Wang
\IEEEcompsocitemizethanks{\IEEEcompsocthanksitem Xuan Dong, Jiangtao
(Gene) Wen and Yi (Amy) Pang are with the Computer Science Department,
 Tsinghua University, Beijing, China, 100084.
 Weixin Li and Guan Wang are with the School of Computer Science and Engineering,
  Beihang University, Beijing, China, 100191.\protect\\
E-mail: jtwen@tsinghua.edu.cn
\IEEEcompsocthanksitem}
\thanks{}}

\markboth{}
{Shell \MakeLowercase{\textit{et al.}}: Bare Demo of IEEEtran.cls
for Journals}

 \maketitle

\begin{abstract}
\emph{We describe a novel integrated algorithm for real-time
enhancement of video acquired under challenging lighting conditions.
Such conditions include low lighting, haze, and high dynamic range
situations. The algorithm automatically detects the dominate
source of impairment, then depending on whether it is low lighting,
haze or others, a corresponding pre-processing is applied to the
input video, followed by the core enhancement algorithm. Temporal
and spatial redundancies in the video input are utilized to
facilitate real-time processing and to improve temporal and spatial
consistency of the output. The proposed algorithm can be
used as an independent module, or be integrated in either a
video encoder or a video decoder for further optimizations.}
\end{abstract}
\IEEEpeerreviewmaketitle

\section{Introduction}
As video surveillance equipments and mobile devices such as digital
cameras, smart phones and netbooks are increasingly widely deployed,
cameras are expected to acquire, record and sometimes compress and
transmit video content in all lighting and weather conditions. The
majority of cameras, however, are not specifically designed to be
all-purpose and weather-proof, rendering the video footage
unusable for critical applications under many circumstances.

Image and video processing and enhancement including gamma
correction, de-hazing, de-bluring and etc. are well-studied areas
with many successful algorithms proposed over the years. Although
different algorithms perform well for different lighting
impairments, they often require tedious and sometimes manual
input-dependent fine-tuning of algorithm parameters. In addition,
different specific types of impairments often require different specific
algorithms.

Take the enhancement of videos acquired under low lighting
conditions as an example. To mitigate the problem, far and near
infrared based techniques (\cite{FIR1,FIR2,NIR1,NIR2}) are used in
many systems, and at the same time, various image processing based
approaches have also been proposed.
 Although far and near infrared systems are useful for
detecting objects such as pedestrians and animals in low lighting
environments, especially in ``professional'' video surveillance
systems, they suffer from the common disadvantage that detectable
objects must have a temperature that is higher than their surroundings. In
many cases where the critical object has a  temperature similar to
its surroundings, e.g. a big hole in the road, the infrared systems
are not as helpful. Furthermore, infrared systems are usually more
expensive, harder to maintain, with a relatively shorter life-span
than conventional systems. They also introduce extra, and often
times considerable power consumption. In many consumer applications such as video capture and
communications on smart phones, it is usually not feasible to deploy
infrared systems due to such cost and power consumption issues.
Conventional low lighting image and video processing enhancement algorithms
such as  \cite{ICCV07} and \cite{Siggraph05} often work by reducing noise in the input low lighting video
followed by contrast enhancement techniques such as
tone-mapping, histogram stretching and equalization, and gamma
correction to recover visual information in low lighting images and videos. Although these algorithms can
lead to very visually pleasing enhancement results, they are usually
too complicated for practical real-time applications, especially on
mobile devices. For example, the processing speed of the algorithm
in \cite{ICCV07} was only 6 fps even with GPU acceleration. In
\cite{Siggraph05}, recovering each single image required more than one
minute.

In this paper, we describe a novel integrated video enhancement
algorithm applicable to a wide range of input impairments. It has
low computational and memory complexities that are both within the realm of reasonable
availability of many mobile devices. In our system, a low complexity
automatic module first determines the pre-dominate source of impairment
in the input video. The input is then pre-processed based on the particular
source of impairment, followed by processing by the core enhancement
module. Finally, post-processing is applied to produce the enhanced
output. In addition, spatial and temporal correlations
were utilized to improve the speed of the algorithm and visual quality of the output, enabling it
to be embedded into video encoders or decoders to share
temporal and spatial prediction modules in the video codec
to further lower complexity.

The paper is organized as the following. In Section \ref{sec:algo},
we present the heuristic evidences that motivated the
idea in this paper. In Section \ref{sec:enh}, we
explain the core enhancement algorithm in detail, while in
Section \ref{sec:opt} we describe various algorithms for reducing the
computational and memory complexities. Sections \ref{sec:results}
contains the experimental results. Given that in real-world
applications, the video enhancement module could be
deployed in multiple stages of the end to end procedure, e.g. before
compression and transmission/storage, or after compression and
transmission/storage but before decompression, or after
decompression and before the video content displayed on the monitor,
we examine the complexity and RD tradeoff associated with applying
the proposed algorithm in these different steps in the experiments.
Finally we conclude the paper and future works in Section
\ref{sec:conclusions}.

\section{A Novel Integrated Algorithm for Video Enhancement} \label{sec:algo}

\begin{figure}[h]
\centering
\includegraphics[width=0.48\textwidth,height=0.3\textwidth]{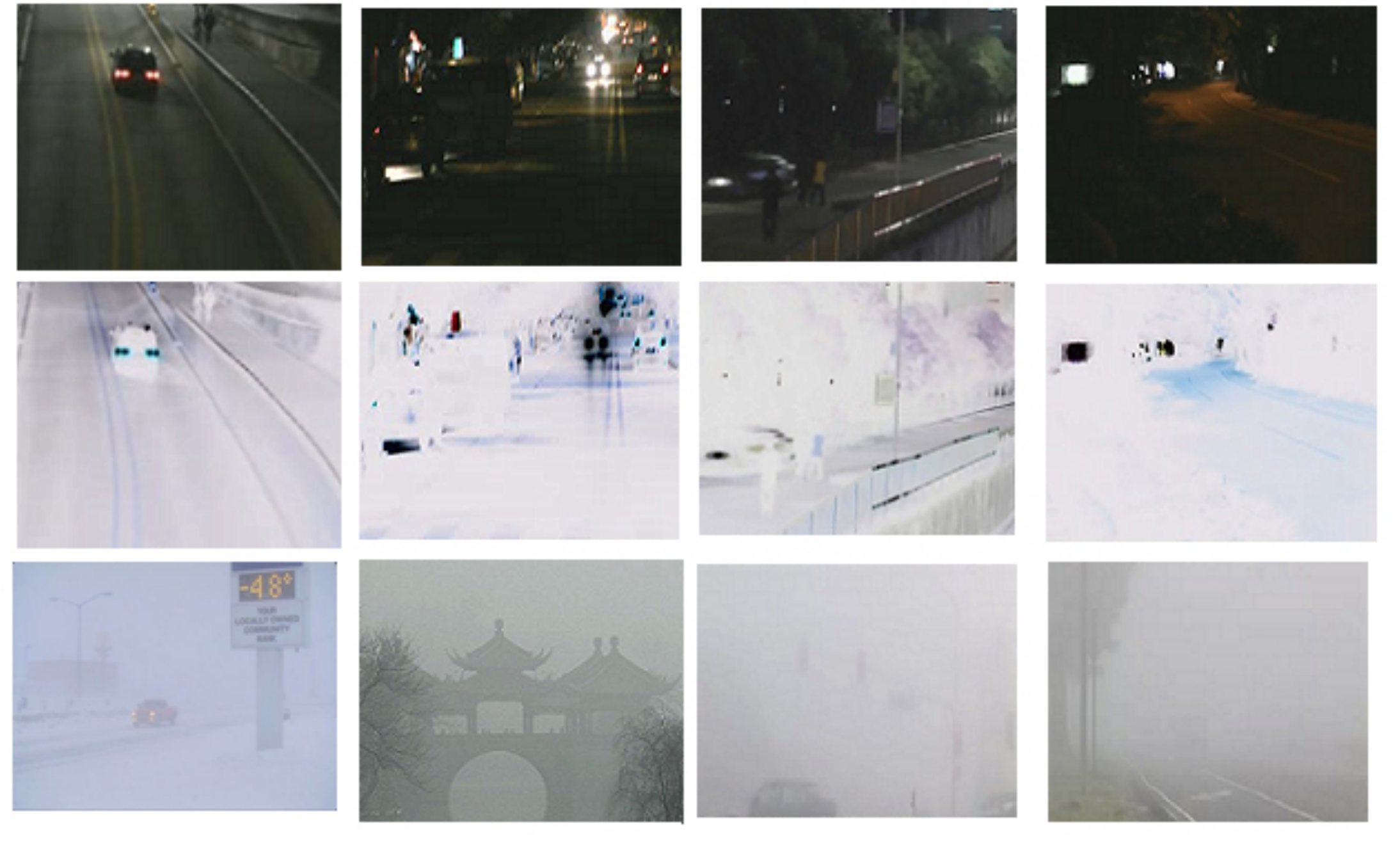}
   \caption{Examples of original (Top), inverted low lighting videos/images (Middle) and haze videos/images (Bottom).}
   \label {fig:day_night}
\end{figure}

The motivation for our algorithm is a key observation that if we
perform a pixel-wise inversion of low lighting videos or high
dynamic range videos, the results look quite similar to hazy videos.
As an illustrative example, we randomly selected (by Google) and
captured a total of 100 images and video clips in haze, low lighting
and high dynamic range weather conditions respectively. Some
examples are shown in Fig. \ref{fig:day_night}. Here, the
``inversion'' operation is simply
\begin {equation}
R^c (x) = 255 - I^c (x), \label{eq:invert}
\end{equation}
where $R^c(x)$ and $I^c(x)$ are intensities for the corresponding
color (RGB) channel $c$ for pixel $x$ in the input and inverted frame respectively.

As can be clearly seen from Fig. \ref{fig:day_night}, at least
visually, the video in hazy weather are similar to the inverted output
of videos captured in low lighting and high dynamic range
conditions. This is intuitive because as illustrated in \cite{1924},
in all these cases, e.g. hazy videos and low lighting
videos, light captured by the camera is blended with the airlight
(ambient light reflected into the line of sight by atmospheric
particles). The only difference is the actual brightness of the
airlight, white in the case of haze videos, black in the case of low
lighting and high dynamic range videos.

The observation is confirmed by various haze detection algorithms.
We implemented haze detection using
the HVS threshold range based method \cite{R.Lim}, the
Dark Object Subtraction (DOS) approach \cite{Song}, and the spatial
frequency based technique \cite{Yong}, and found that
hazy, inverted low lighting videos and inverted high dynamic
range videos were all classified as hazy video clips, as opposed to
``normal'' clips.

\begin{figure}
\centering
\includegraphics[width=0.38\textwidth,height=0.3\textwidth]{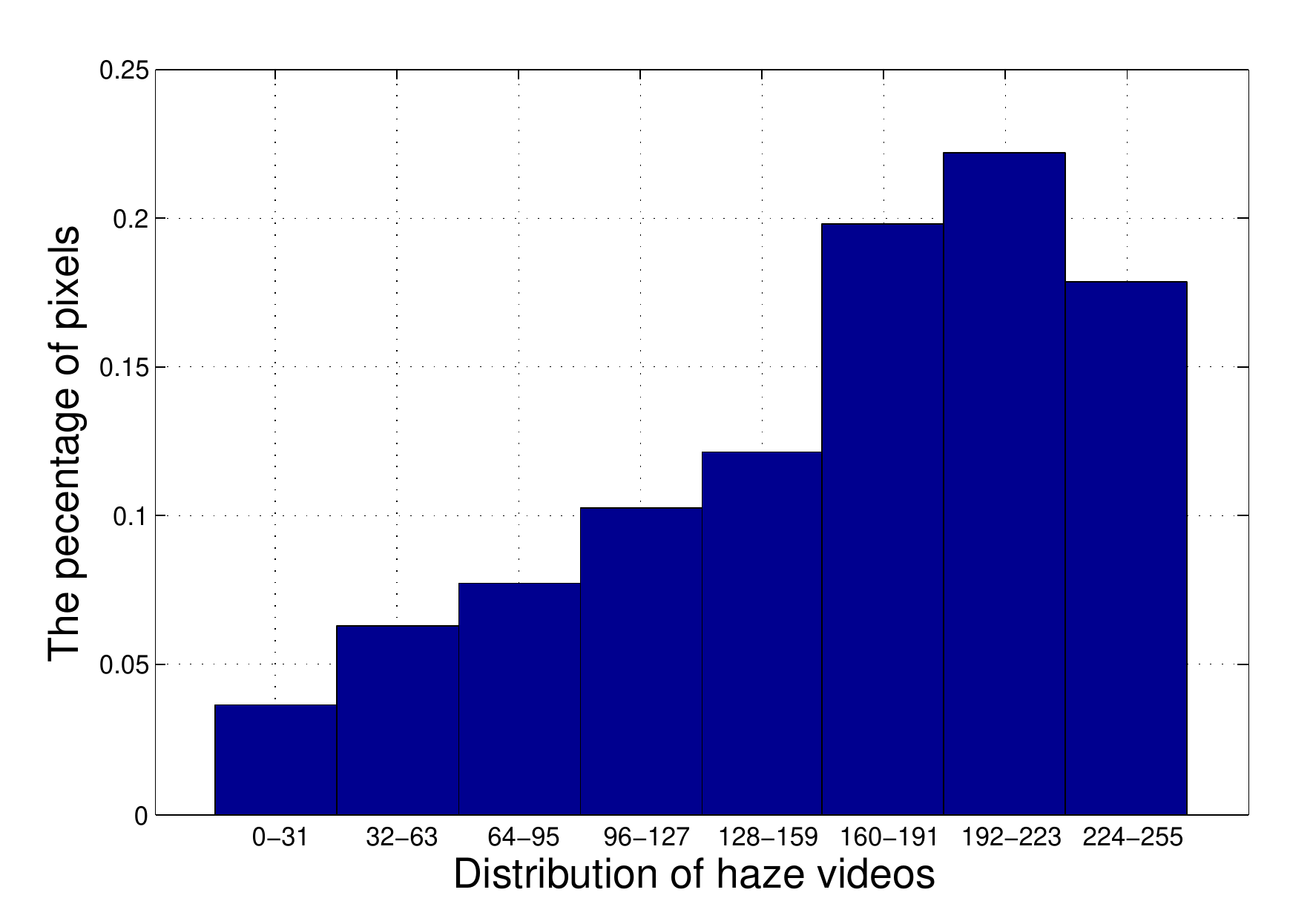}\\
\includegraphics[width=0.38\textwidth,height=0.3\textwidth]{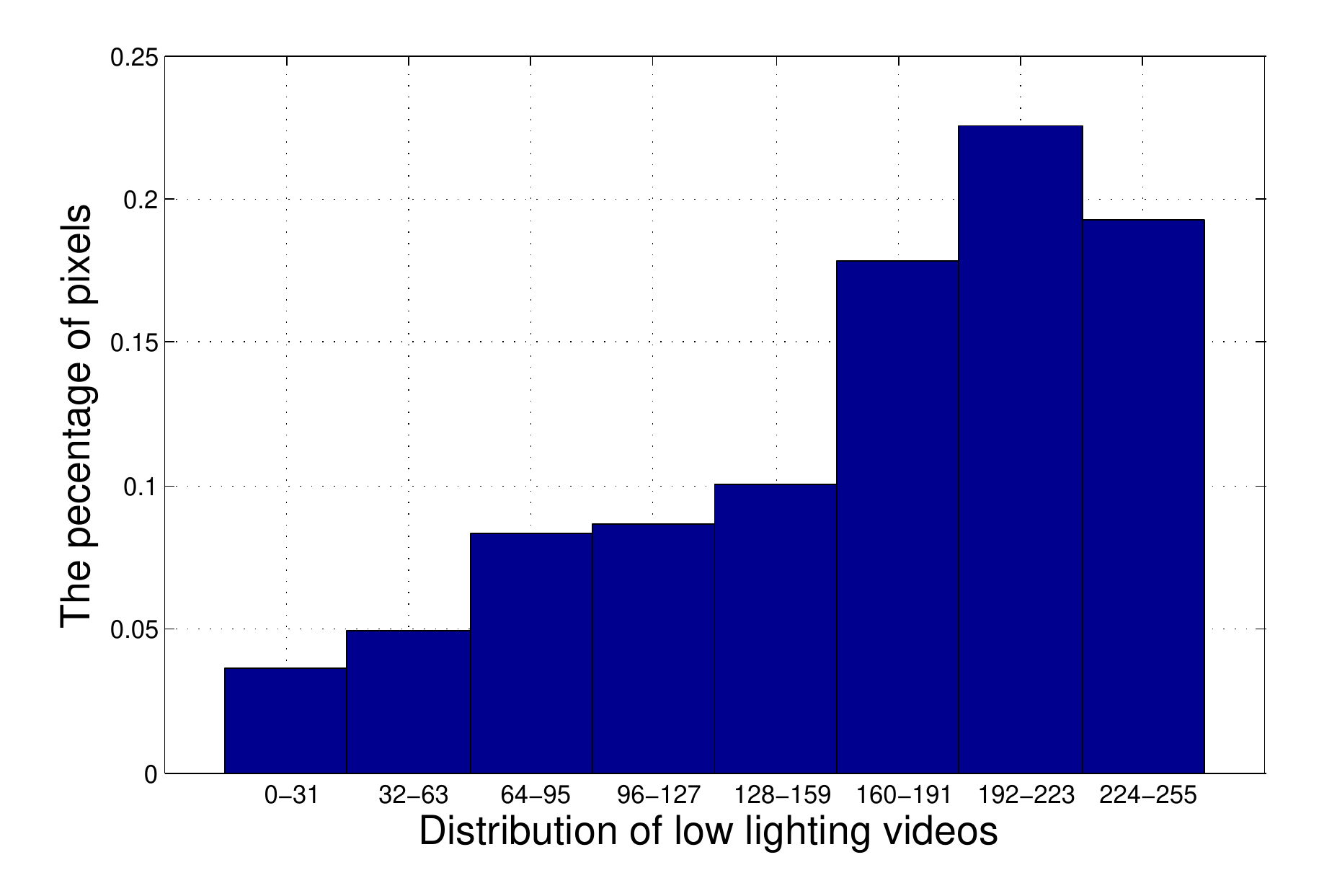}\\
\includegraphics[width=0.38\textwidth,height=0.3\textwidth]{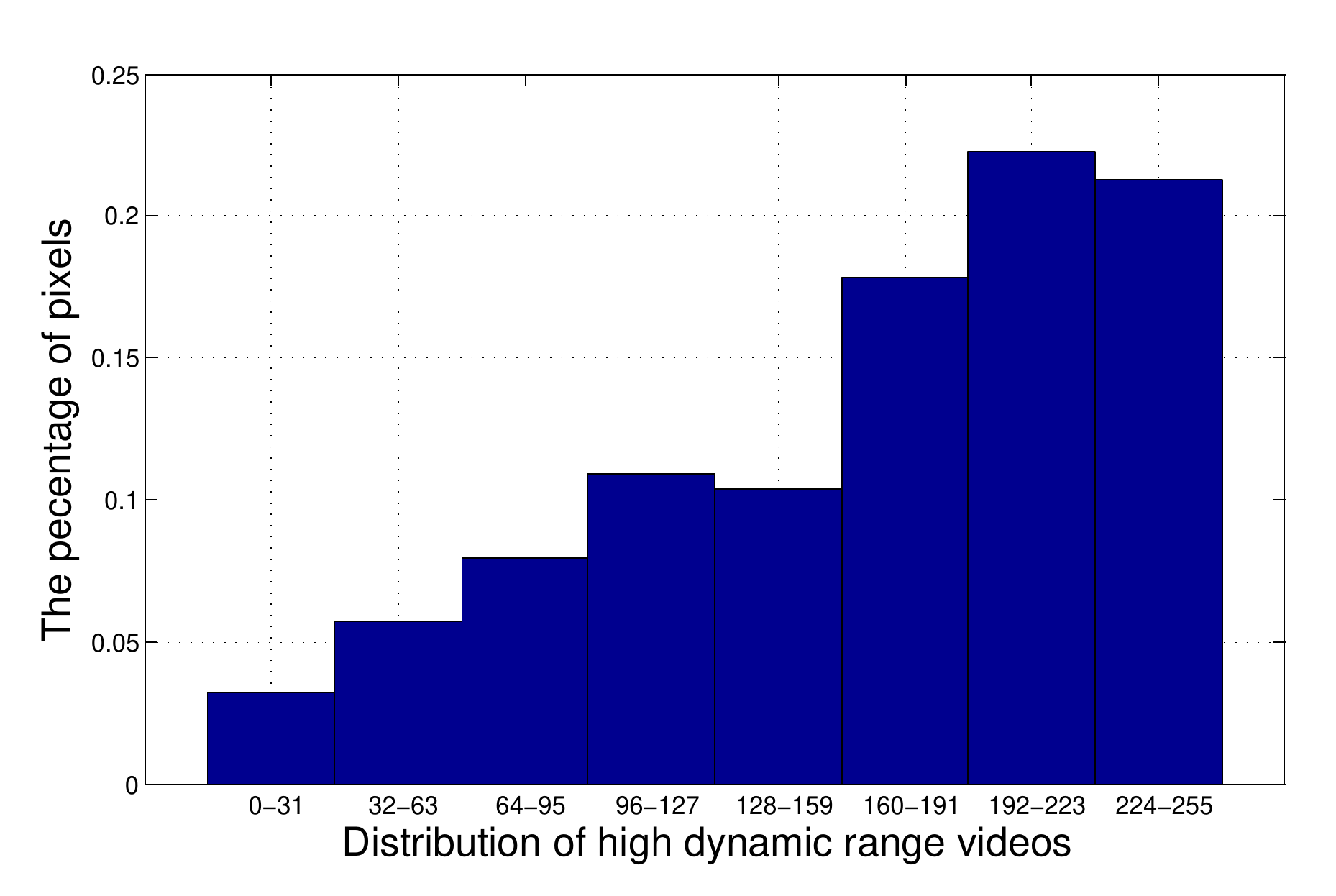}\\
   \caption{The histogram of the minimum intensity of each pixel's three color channels of haze videos (Top), low lighting videos (Middle) and high dynamic range videos (Bottom).}
   \label{fig:chi_square}
\end{figure}

\begin{table*}
\caption {Results of chi square tests} \centering
\begin{tabular}{|c|c|c|}
  \hline
    Data of chi square test& Degrees of Freedom & Chi square values \\ \hline
  Haze videos and inverted low lighting videos & 7 & 13.21 \\ \hline
  Haze videos and inverted high dynamic range videos & 7 & 11.53 \\
  \hline

\end{tabular}
\label{table:chi-square}
\end{table*}

We also performed the chi-square test to examine the statistical
similarities between hazy videos and inverted low lighting and high
dynamic range videos. The chi-square test is a standard statistical
tool widely used to determine if observed data are consistent with a
specific hypothesis. As explained in \cite{Chi}, in chi-square
tests, a $p$ value is calculated, and usually, if $p > 0.05$, it is
reasonable to assume that the deviation of the observed data from
the expectation is due to chance alone. In our experiments, the
expected distribution was calculated from hazy videos and the
observed statistics from inverted low lighting and high dynamic
range videos were tested. In the experiments, we divided the range
[0, 255] of color channel intensities into eight equal intervals,
corresponding to a degree of freedom of 7. According to the
chi-square distribution table, if we adopt the common standard of $
p> 0.05 $, the corresponding upper threshold for the chi-square
value should be 14.07. The histogram of the minimum intensities of
all color channels of all pixels for hazy videos, inverted low
lighting and inverted high dynamic range videos were used in the
tests, some examples are shown in Fig. \ref{fig:chi_square}. The
results of the chi-square tests are given in Table
\ref{table:chi-square}. As can be seen from the table, the
chi-square values are far smaller than 14.07, demonstrating that our
hypothesis of the similarities between haze videos and inverted low
lighting videos, and between haze videos and high dynamic range
videos is reasonable.

\begin{figure}
\centering
\includegraphics[width=0.24\textwidth,height=0.21\textwidth]{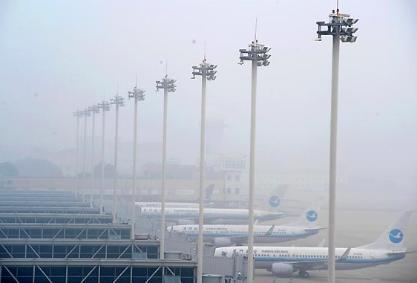}
\includegraphics[width=0.24\textwidth,height=0.21\textwidth]{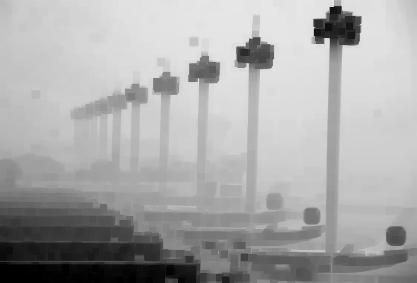}
   \caption{Examples of hazy videos/images (Left) and their dark channel images (Right).}
   \label{fig:haze_jdark}
\end{figure}

\begin{figure}
\centering
\includegraphics[width=0.24\textwidth,height=0.21\textwidth]{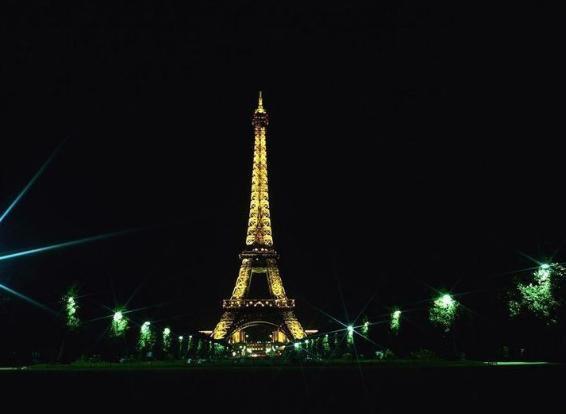}
\includegraphics[width=0.24\textwidth,height=0.21\textwidth]{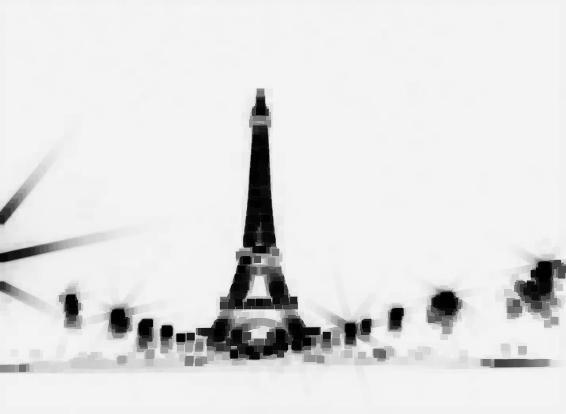}
   \caption{Examples of low lighting videos/images (Left) and their dark channel images (Right).}
   \label{fig:low_lighting_jdark}
\end{figure}

\begin{figure}
\centering
\includegraphics[width=0.24\textwidth,height=0.21\textwidth]{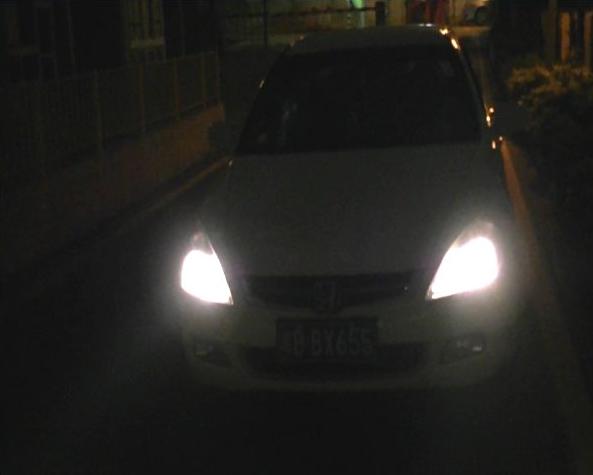}
\includegraphics[width=0.24\textwidth,height=0.21\textwidth]{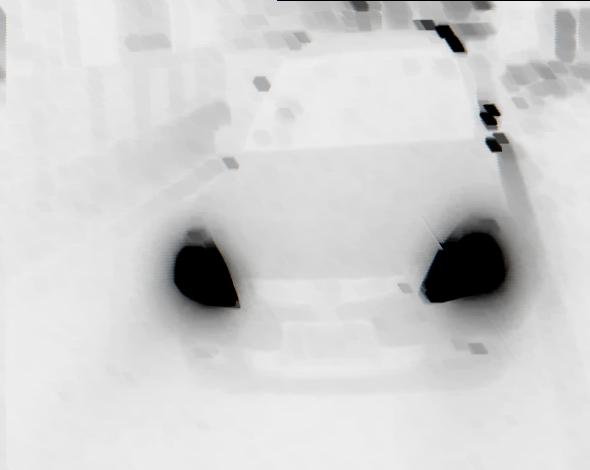}
   \caption{Examples of high dynamic range videos/images (Left) and their dark channel images (Right).}
   \label{fig:high_lighting_jdark}
\end{figure}

Through the experiments, we also found that the pixels whose minimum
intensity of the three color channels was low had a very high
probability of locating in regions of houses, vehicles and etc.. We
introduce the concept of Region of Interests (ROIs) for these
regions. To visually demonstrate the ROIs, we calculated the image
of minimum intensities of color channels for hazy videos, inverted
low lighting videos and inverted high dynamic range videos. Three
examples are shown in Fig. \ref{fig:haze_jdark}, Fig.
\ref{fig:low_lighting_jdark} and Fig. \ref{fig:high_lighting_jdark}.

In conclusion, through visual observation and statistical tests, we found that
video captured in a number of challenging lighting conditions is statistically
and visually similar to hazy videos. Therefore, it is
conceivable that a generic core module could be used for the
enhancement of all these cases.

\section{A Generic Video Enhancement Algorithm Based on Image De-Hazing } \label{sec:enh}

\subsection{Core De-Hazing Based Enhancement Module}

Since after proper pre-processing, videos captured in challenging lighting
conditions, e.g. low lighting and high dynamic range, exhibit strong
similarities with hazy videos in both the visual and statistical
domains, the core enhancement algorithm in our proposed system is an
improved de-hazing algorithm based on \cite{He}.

As mentioned above, most of the existing advanced haze-removal
algorithms (\cite{Fattal},
 \cite{He}, \cite{Tan}, and \cite{Na}) are based on the well-known degradation model proposed by
Koschmieder in 1924 \cite{1924}:
\begin{equation}
R(x) = J(x)t(x) + A(1 - t(x)), \label{eqn:R_x}
\end{equation}
 where $A$ is the global airlight, $R(x)$ is the intensity of pixel $x$
that the camera captures, $J(x)$ is the intensity of the original objects or
 scene, and $t(x)$ is the medium transmission function describing the percentage
of the light emitted from the objects or scene that reaches the
camera. This model assumes that each degraded pixel is a combination
of the airlight and the unknown surface radiance. The medium
transmission describes what percentage of the light emitted from the
objects or scene can reach the camera. And it is determined by the
scene depth and the scattering coefficient of the atmosphere. For
the same video where the scattering coefficient of the atmosphere is
constant, the light is more heavily affected by the airlight in sky
regions because of the longer distance. In other regions such as
vehicles, houses and etc., especially those nearby, the light is
less affected by the airlight.

The critical part of all the algorithms based on the Koschmieder model is to
estimate $A$ and $t(x)$ from the recoded image intensity $I(x)$ so
as to recover the $J(x)$ from $I(x)$. For example, in \cite{Fattal},
Independent Component Analysis is used to estimate the medium
transmission and the airlight. In \cite{He}, the medium transmission
and airlight are estimated by the Dark Channel method, based on the
assumption that the medium transmission in a local patch is
constant.

In our system, we estimate $t(x)$ according to \cite{He} using
\begin{equation}
t(x) = 1 -\omega \mathop {\min }\limits_{c \in \{ r,g,b\} } \left \{
\mathop {\min }\limits_{y \in \Omega (x)} \frac{{R^c (y)}}{{A^c }}
\right \},
\end{equation}
where $\omega=0.8$ and $\Omega (x)$ is a local $9\times9$ block
centered at $x$ in this paper. As our system also targets
application in mobile devices, the cpu-and-memory-costly soft
matting method proposed in \cite{He} is not implemented in our
algorithm.
\begin{figure}
\includegraphics[width=0.24\textwidth]{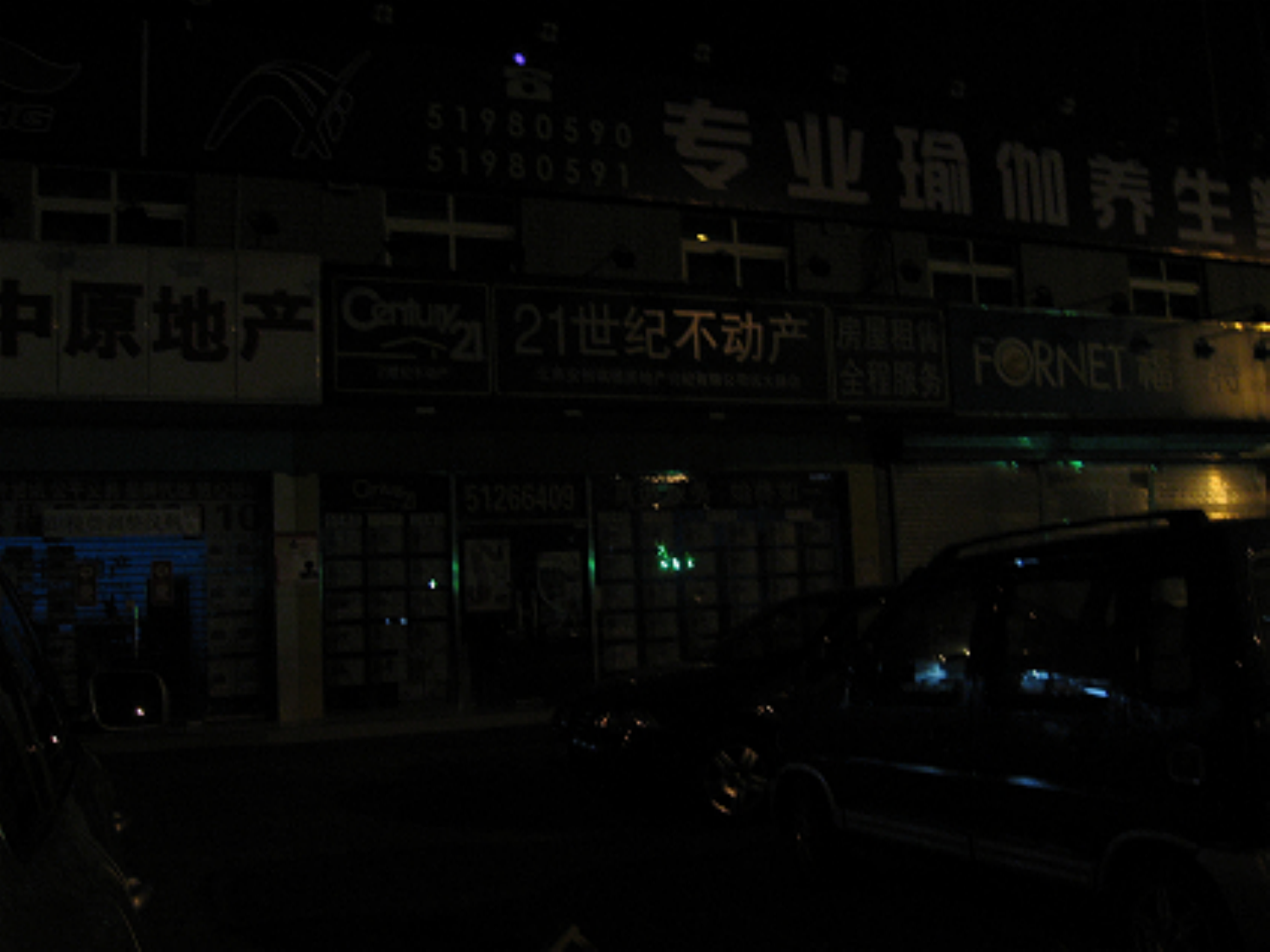}
\includegraphics[width=0.24\textwidth]{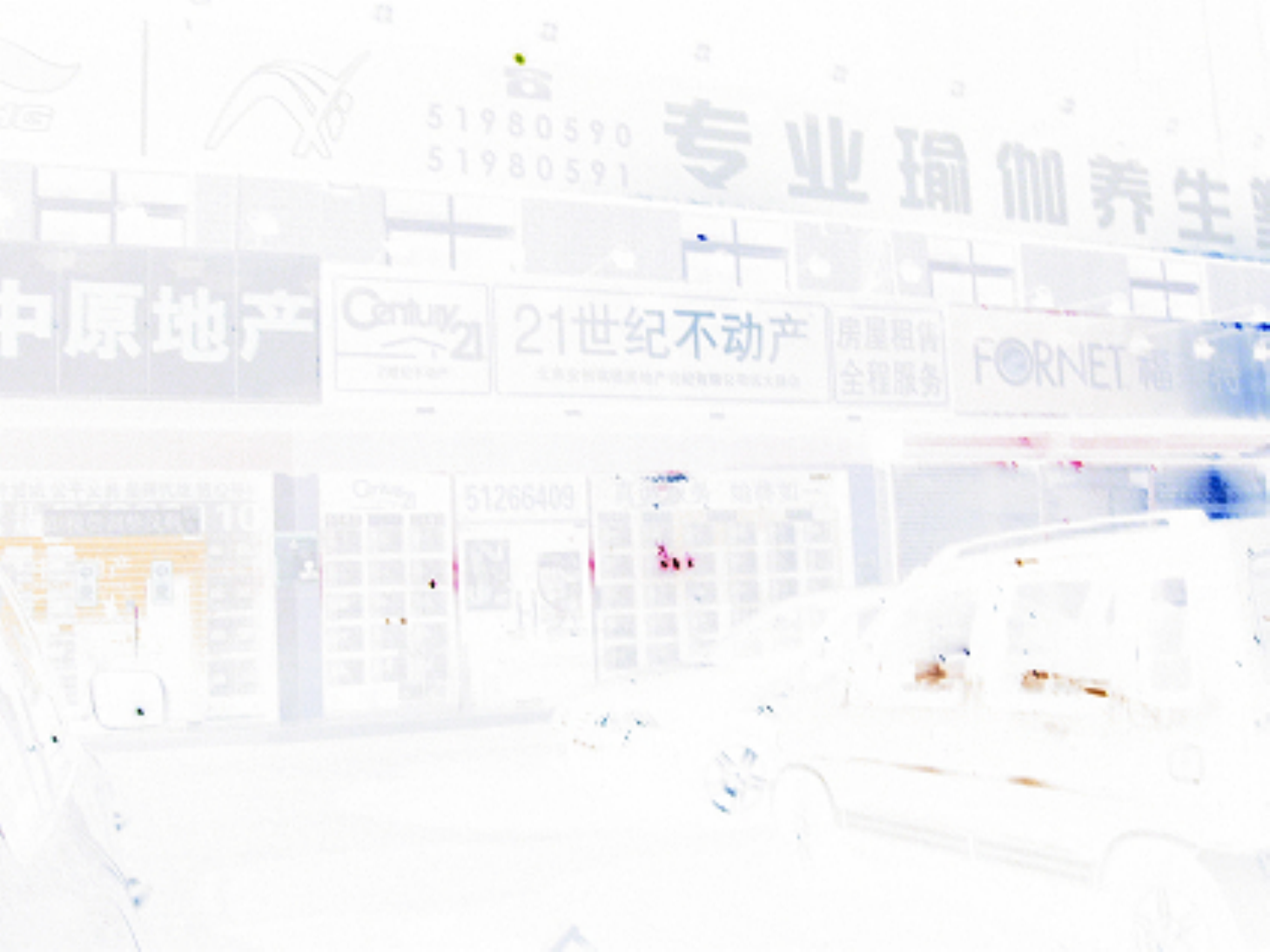}
\includegraphics[width=0.35\textwidth,height=0.01\textwidth]{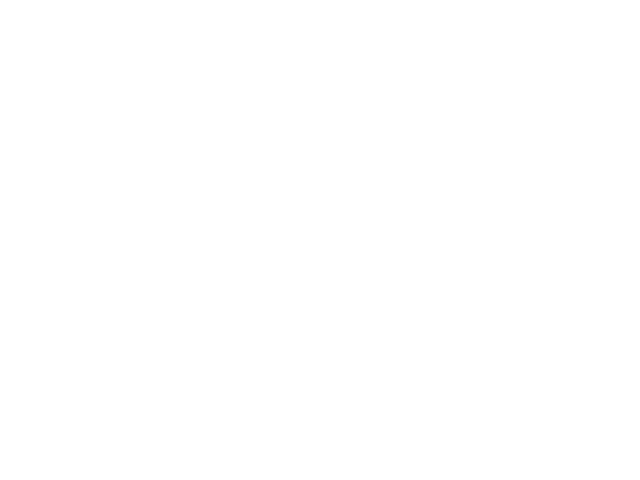}
\includegraphics[width=0.24\textwidth]{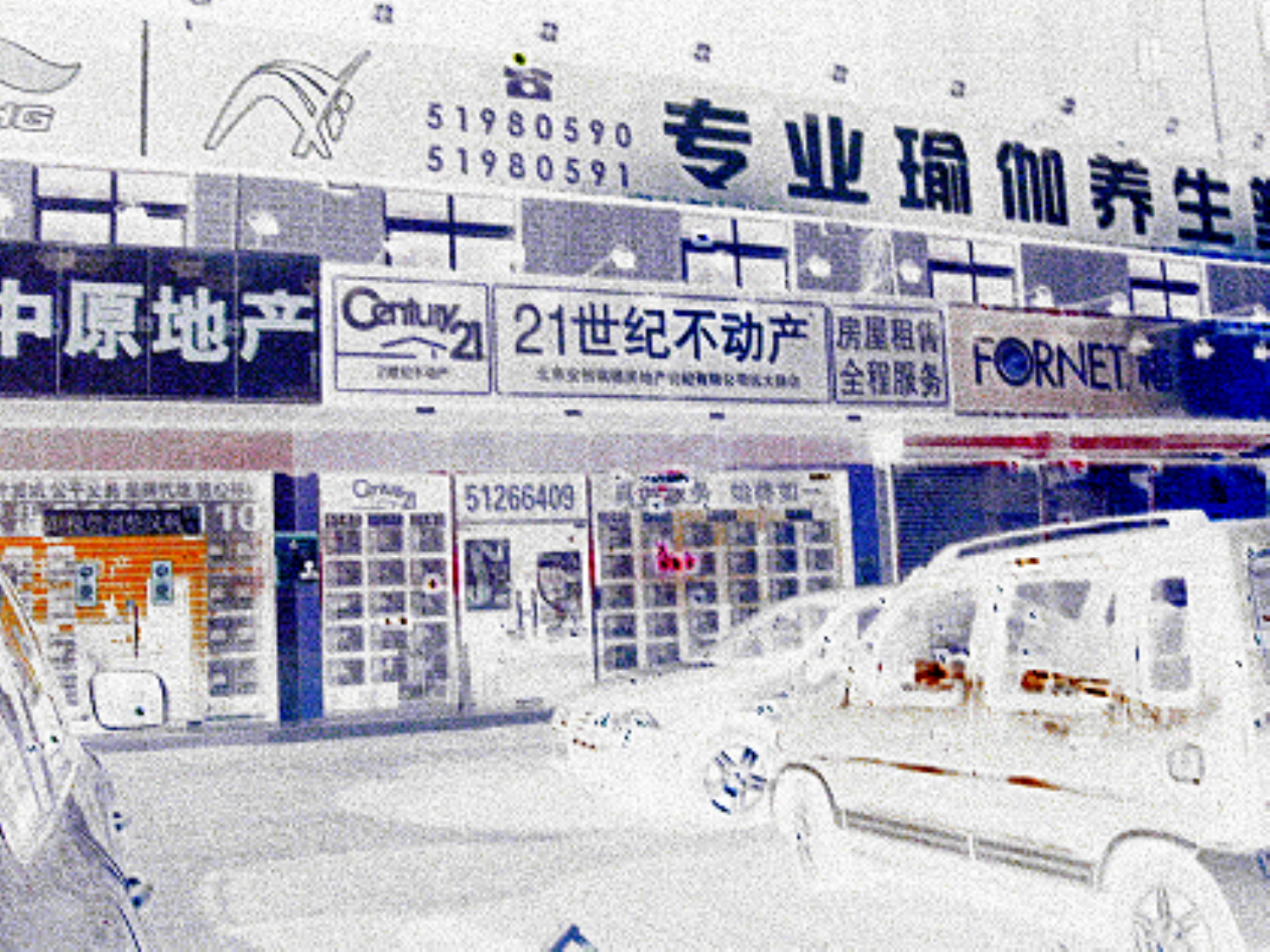}
\includegraphics[width=0.24\textwidth]{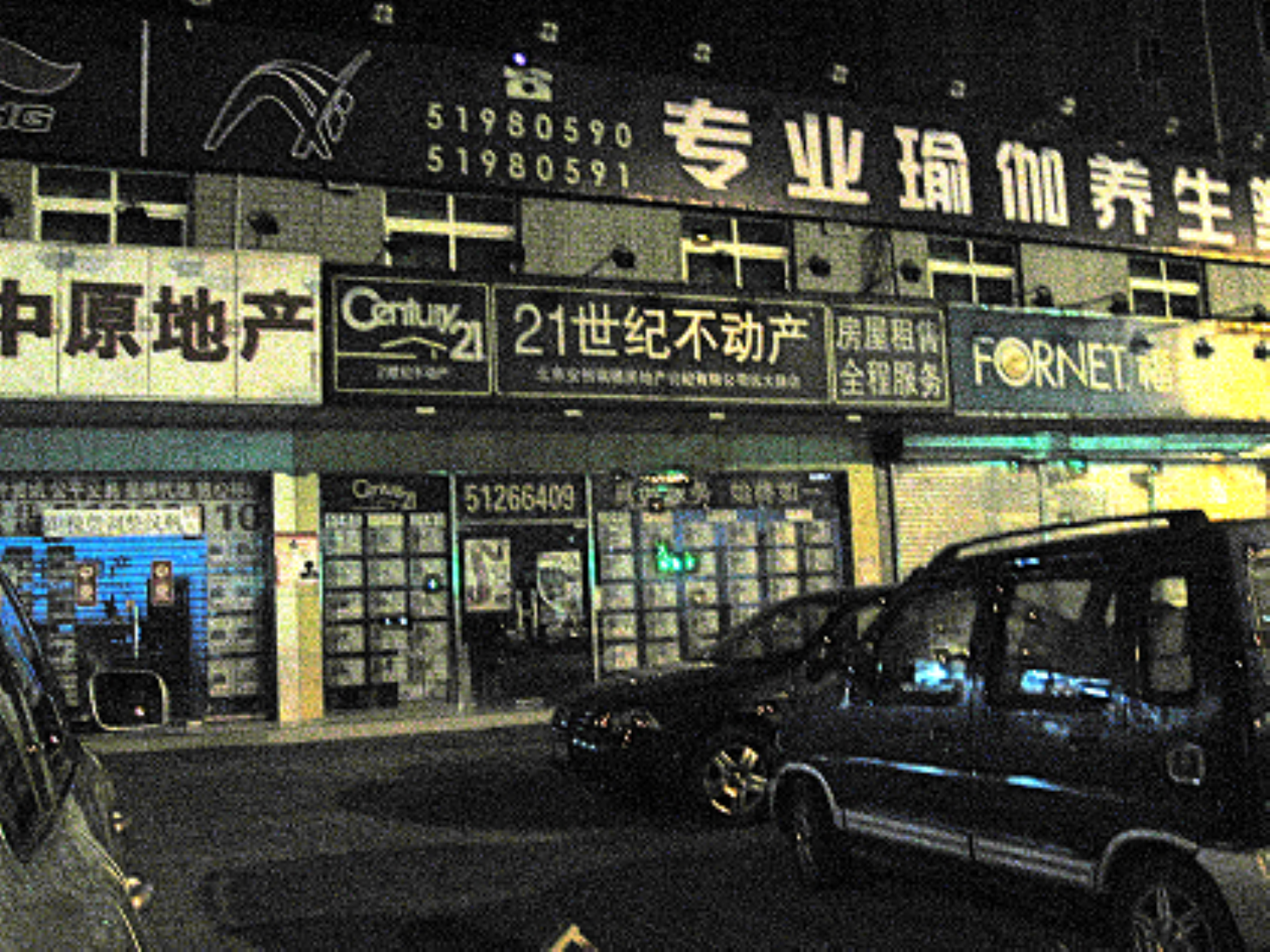}
   \caption{Examples of processing steps of low lighting enhancement algorithm: input image $I$ (Top left), inverted input image $R$ (Top right), haze removal result $J$ of the image $R$ (Bottom left), and output image (Bottom right).}
   \label {fig:A_car}
\end{figure}

\begin{figure*}
\centering
\includegraphics[width=0.18\textwidth,height=0.11\textwidth]{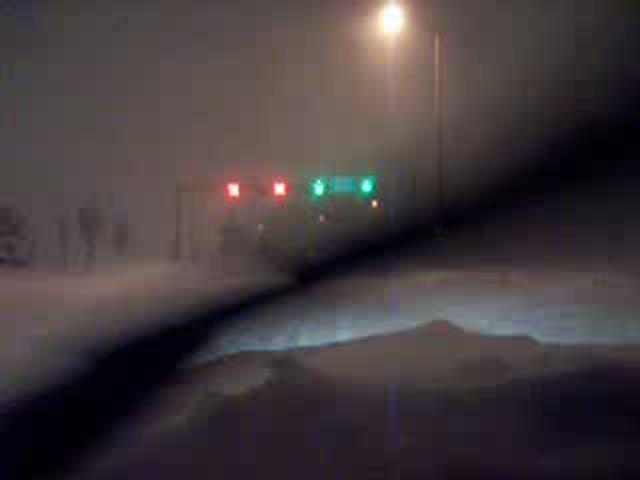}
\includegraphics[width=0.18\textwidth,height=0.11\textwidth]{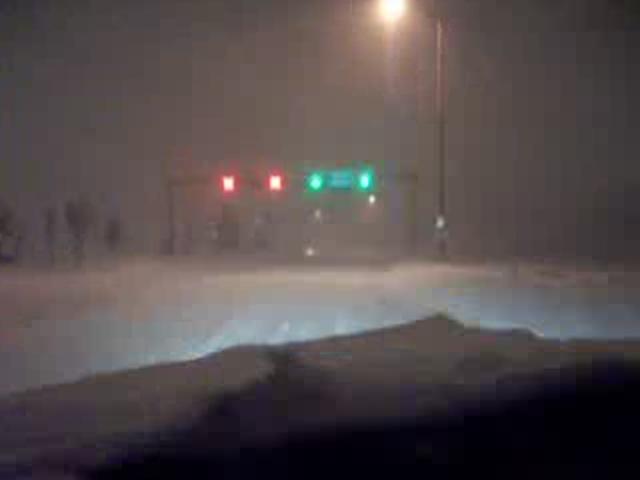}
\includegraphics[width=0.18\textwidth,height=0.11\textwidth]{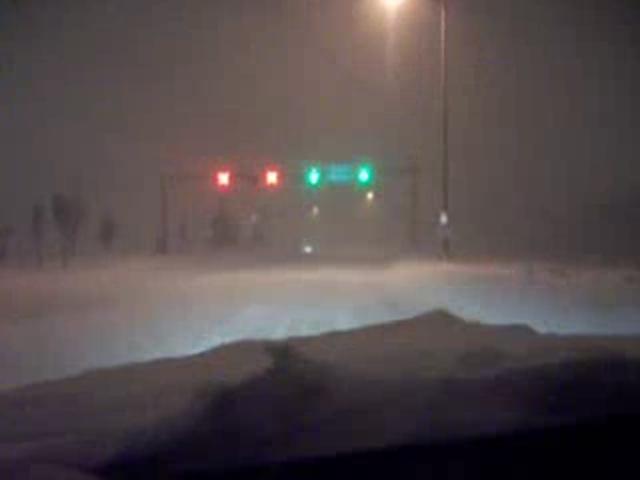}
\includegraphics[width=0.18\textwidth,height=0.11\textwidth]{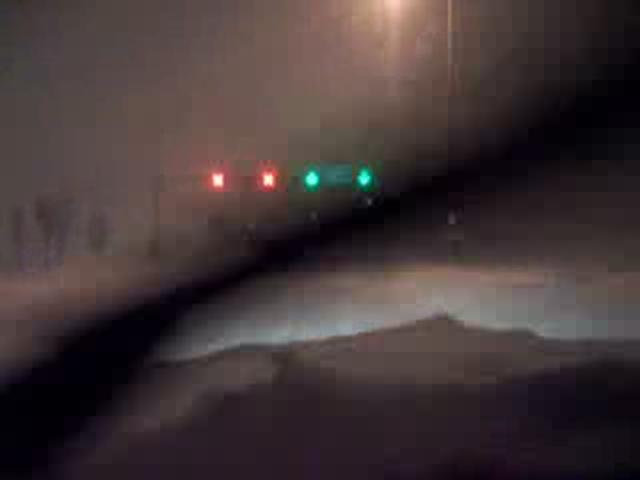}
\includegraphics[width=0.18\textwidth,height=0.11\textwidth]{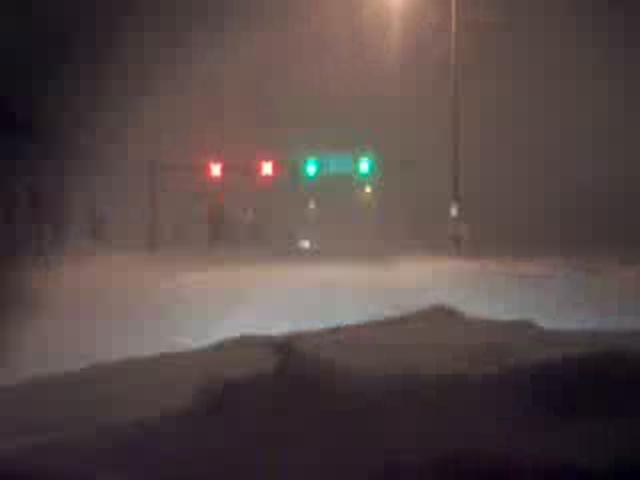}
\includegraphics[width=0.9\textwidth,height=0.01\textwidth]{blank.jpg}
\includegraphics[width=0.18\textwidth,height=0.11\textwidth]{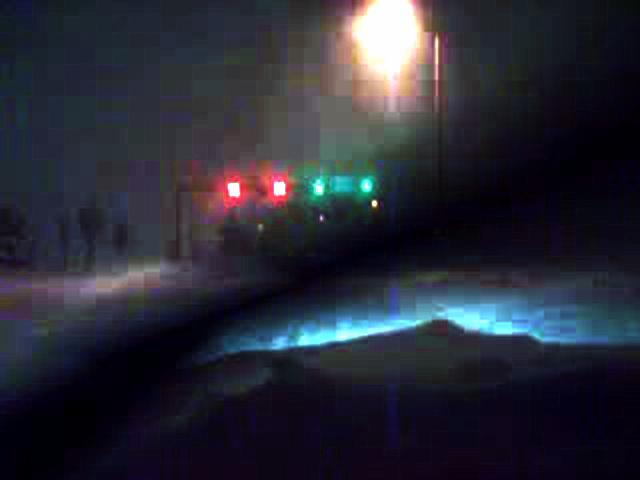}
\includegraphics[width=0.18\textwidth,height=0.11\textwidth]{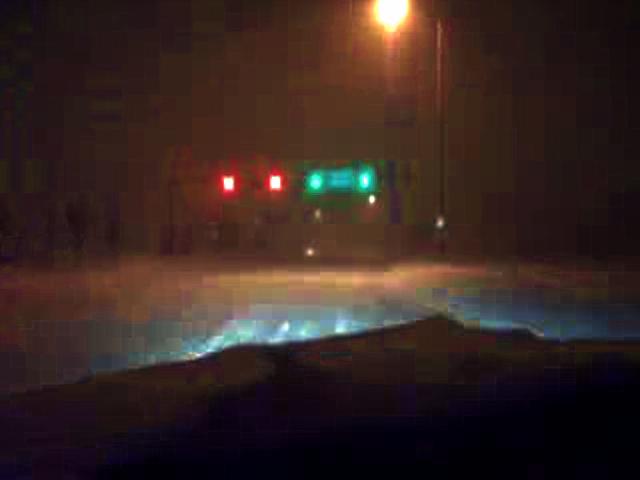}
\includegraphics[width=0.18\textwidth,height=0.11\textwidth]{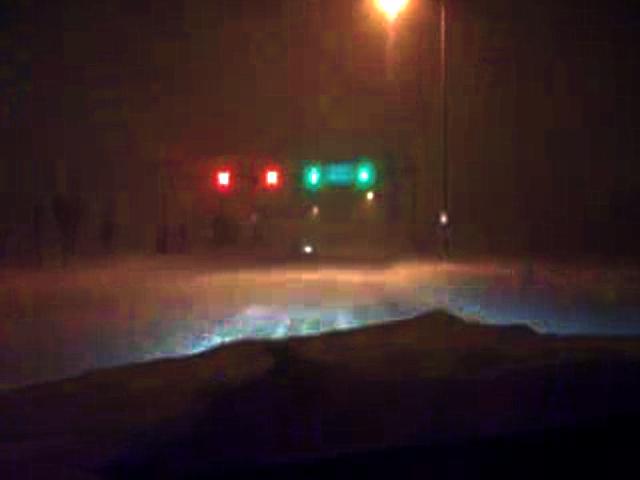}
\includegraphics[width=0.18\textwidth,height=0.11\textwidth]{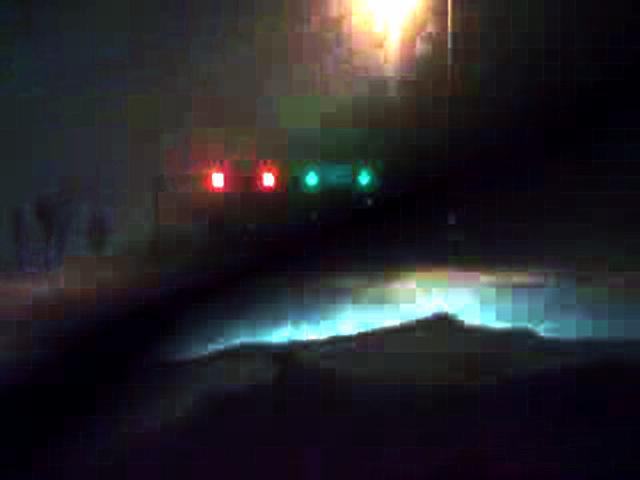}
\includegraphics[width=0.18\textwidth,height=0.11\textwidth]{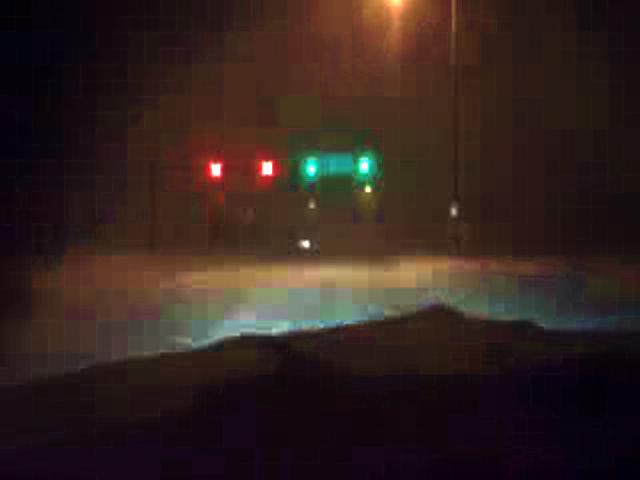}
\includegraphics[width=0.9\textwidth,height=0.01\textwidth]{blank.jpg}
\includegraphics[width=0.18\textwidth,height=0.11\textwidth]{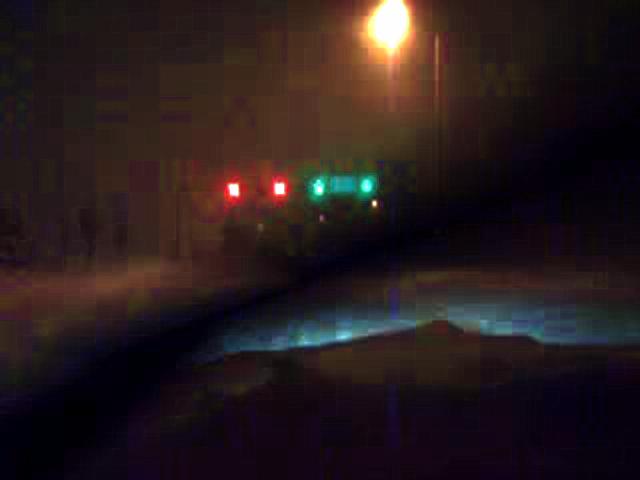}
\includegraphics[width=0.18\textwidth,height=0.11\textwidth]{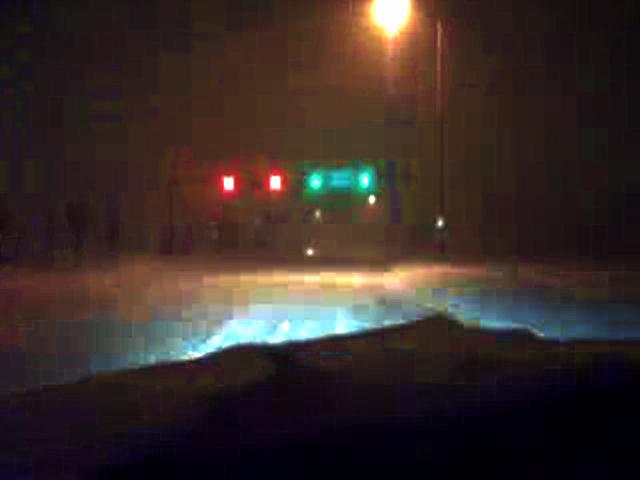}
\includegraphics[width=0.18\textwidth,height=0.11\textwidth]{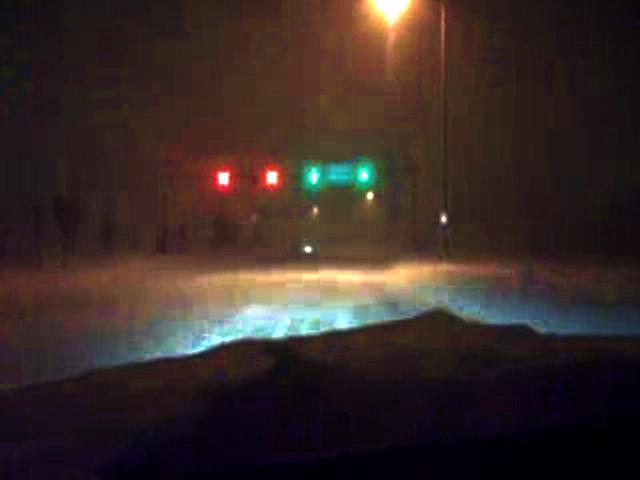}
\includegraphics[width=0.18\textwidth,height=0.11\textwidth]{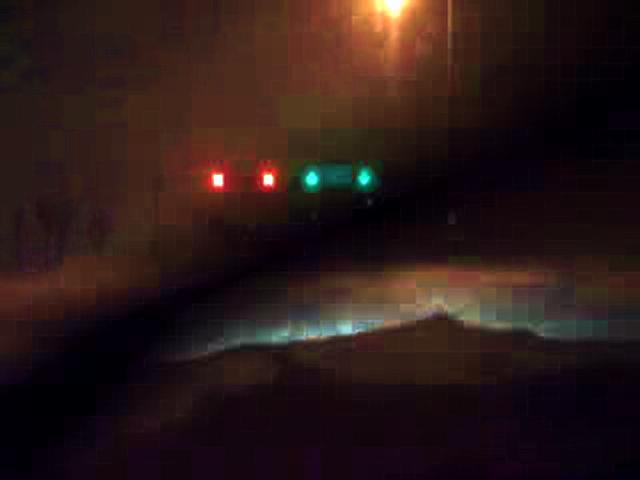}
\includegraphics[width=0.18\textwidth,height=0.11\textwidth]{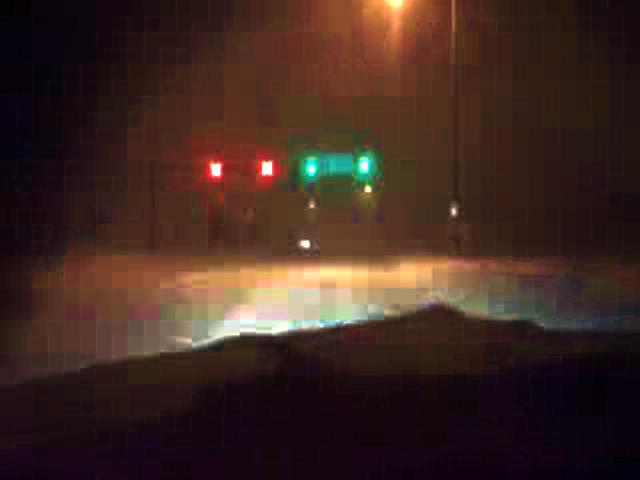}
   \caption{The comparison of original, haze removal, and optimized haze removal video
   clips.
Top: input video sequences. Middle: outputs of image haze removal
algorithm of \cite{He}. Bottom: outputs of haze removal using our
optimized algorithm in calculating airlight.}
   \label {fig:A_shake}
\end{figure*}

To estimate airlight, we first note that the schemes in existing
image haze removal algorithms are usually not robust and even very
small changes to the airlight value might lead to very large changes
to the recovered images or video frames. Therefore, calculating
airlight frame-wise not only increases the overall complexity of the
system, but also introduces visual inconsistency between frames,
thereby creating annoying visual artifacts. Fig. \ref{fig:A_shake}
shows an example using the results of the algorithm in \cite{He}.
Notice the difference between the first and fourth frame in the
middle row.

Based on this observation, we propose to calculate airlight only
once for a Group of Pictures (GOP). This is done for the first
frame of the GOP, then the same value is used for all subsequent frames
in the same GOP. In the implementation, we also incorporated a scene change detection module
so as to detect sudden changes in airlight that are not aligned with
GOP boundaries but merit recalculation.

In our system, to estimate airlight, we first select 100 pixels
whose minimum intensities in all color (RGB) channels are the
highest in the image. Then from these pixels, we choose the single
pixel whose sum of RGB values is the highest. Among successive GOPs,
we refresh the value of airlight using the equation
\begin{equation}
A=A*0.4+A_t*0.6, \label{eqn:A}
\end{equation}
where $A_t$ is the airlight value calculated in this GOP, $A$ is the
global airlight value. This can efficiently avoid severe changes of
the global airlight value $A$, bringing about the excellent
recovered results and saving a large amount of computation at the
same time. Examples of the recovered results are shown in Fig.
\ref{fig:A_shake}. The first and fourth frame in the bottom row change gradually using
our algorithm.

Then, from (\ref{eqn:R_x}), we can find
\begin{equation}
J(x)  = \frac{{R(x) - A}}{{t(x)}} + A \label{eqn:J_x}.
\end{equation}
Although (\ref{eqn:J_x}) works reasonably well for haze removal, through experiments we found
that direct application of equation (\ref{eqn:J_x}) might lead to
under-enhancement for low lighting areas and over-enhancement for
high lighting areas when applied to low lighting video enhancement. To further optimize the calculation of $t(x)$,
we focus on enhancing the ROIs while avoid processing the background,
e.g. sky regions in low lighting and high dynamic range videos. This
not only further reduces computational complexity, but also improves
overall visual quality. To this end, we adjust $t(x)$ adaptively
while maintaining its spatial continuity, so that the resulted video
becomes more smooth visually. We introduce a multiplier $P(x)$ into
equation (\ref{eqn:J_x}), and through extensive experiments, we find
that $P(x)$ can be set as
\begin{equation}
P\left( x \right) = \left\{ {\begin{array}{*{20}c}
   {2t\left( x \right)} & {0 < t\left( x \right) \le 0.5},  \\
   { - 2t^2 \left( x \right) + 8 - \frac{3}{{t\left( x \right)}}} & {0.5 < t\left( x \right) \le 1}.  \\
\end{array}} \right.
\end{equation}
Then (\ref{eqn:J_x}) becomes
\begin{equation}
J(x) = \frac{{R(x) - A}}{{P(x)t(x)}} + A \label{eqn:J_x2}.
\end{equation}

The idea behind (\ref{eqn:J_x2}) is as the following. When $t(x)$ is
smaller than 0.5, which means that the corresponding pixel needs
boosting, we assign $P(x)$ a small value to make $P(x)t(x)$ even
smaller so as to increase the RGB intensities of this pixel. On the
other hand, when $t(x)$ is greater than 0.5, we refrain from overly
boosting the corresponding pixel intensity. When $t(x)$ is close to
1, $P(x)t(x)$ may be larger than 1, resulting in slight ``dulling''
of the pixel, so as to make the overall visual quality more balanced
and pleasant.

\begin{figure}
\centering
\includegraphics[width=0.15\textwidth]{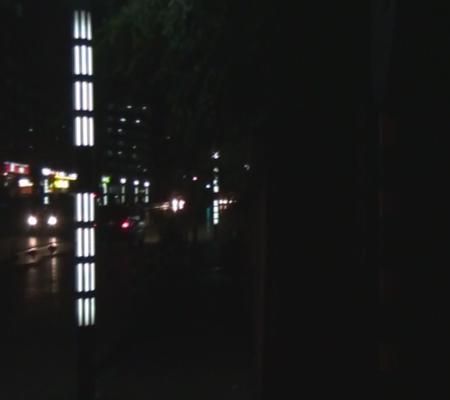}
\includegraphics[width=0.15\textwidth]{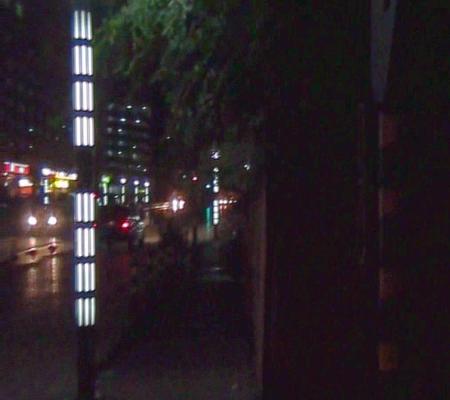}
\includegraphics[width=0.15\textwidth]{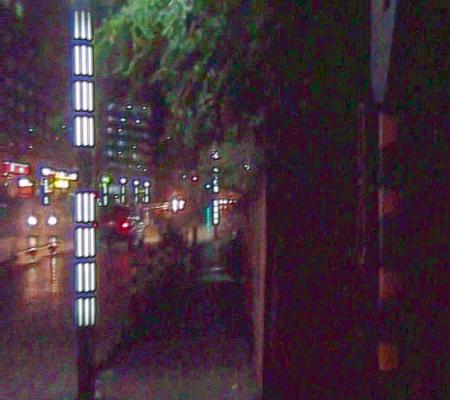}
   \caption{Examples of optimizing low lighting and high dynamic range enhancement algorithm by introducing $P(x)$: Input (Left), output of the enhancement algorithm without introducing $P(x)$ (Middle), and output of the enhancement algorithm by introducing $P(x)$ (Right). }
   \label {fig:Px}
\end{figure}

For low lighting and high dynamic range videos, once $J(x)$ is
recovered, the inversion operation (\ref{eq:invert}) is performed again
to produce the enhanced videos of the original input. This process
is conceptually shown in Fig. \ref{fig:A_car}. The improvement after
introducing $P(x)$ can be seen in Fig. \ref{fig:Px}.

\subsection{Automatic Impairment Source Detection}

\begin{figure}
\centering
\includegraphics[width=0.48\textwidth]{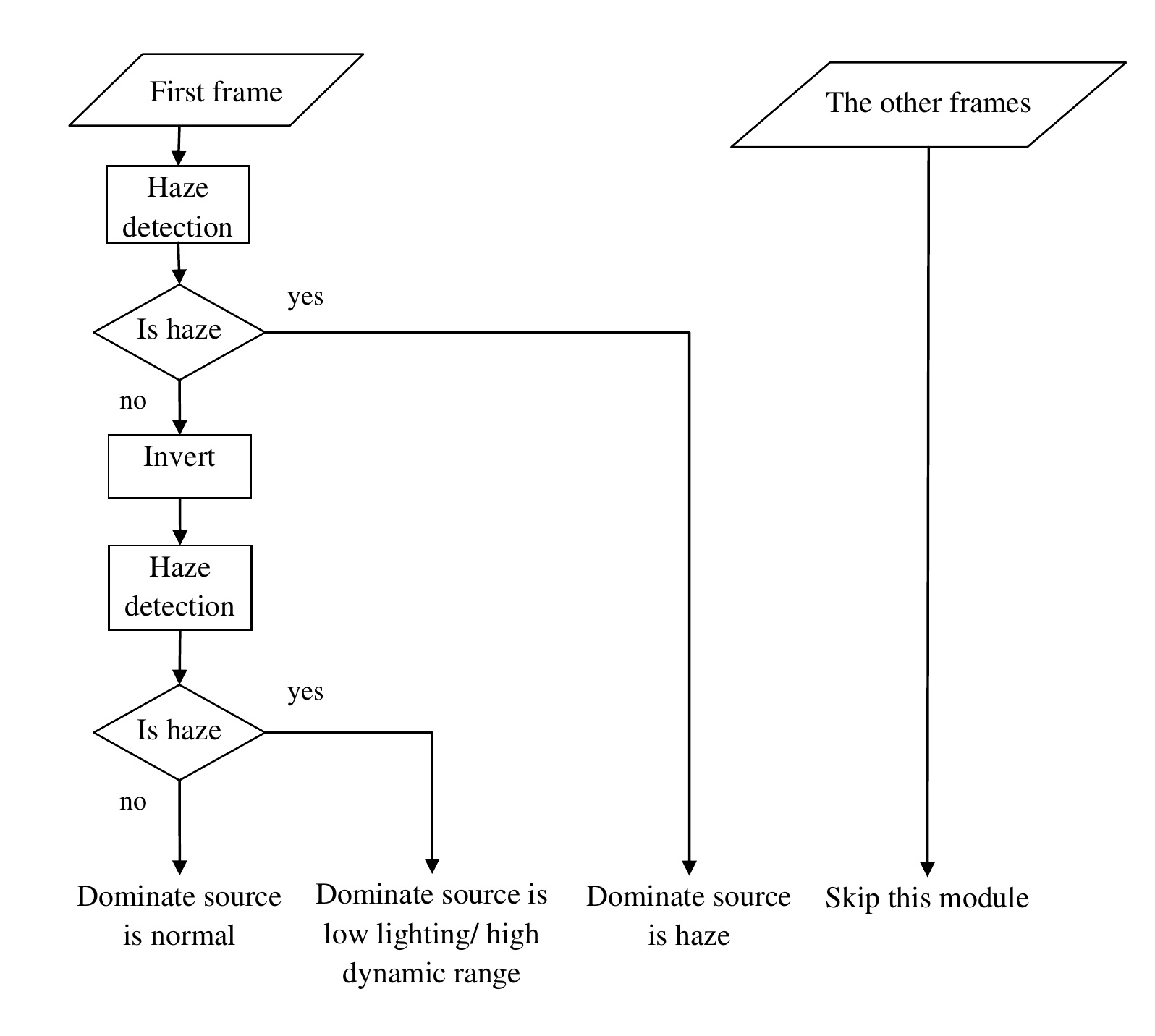}
   \caption{Flow diagram of the module of determining dominate source of impairment.}
   \label {fig:flow_determine}
\end{figure}

As mentioned above, we use the generic video enhancement algorithm
of the previous subsection for enhancing video acquired in a number
of challenging lighting conditions. In addition to this core
enhancement module, the overall system also contains a module for
automatically detecting the main source of visual quality
degradation to determine if the pre-processing by pixel-wise
inversion is required. In the case when pixel-wise inversion is required,
different pixel wise fine tuning may also be introduced so that the eventual output after enhancement is further optimized.
The flow diagram for this automatic detection
system is shown in Fig. \ref{fig:flow_determine}.

Our detection algorithm is based on the technique introduced by R. Lim et al.
\cite{R.Lim}. To reduce complexity, we only perform the
automatic detection for the first frame in a GOP, coupled with a scene change detection.
The corresponding algorithm parameters are shown in Table \ref{table:haze_detection}.
The test is conducted for each pixel in the frame. If the percentage
of hazy pixels in a picture is higher than 60\%, we consider the
picture as a hazy picture. Similarly, if an image is determined to
be a hazy picture after inversion, it is labeled as a low lighting
or high dynamic range image, both of which require the introduction of the
multiplier $P(x)$ into the core enhancement algorithm.
\begin{table}
\caption {Specific parameters of the haze detection algorithm.}
\centering
\begin{tabular}{|c|c|c|}
  \hline
    & Color attribute & Threshold range \\ \hline
  S & $0\sim255$ & $0\sim130$ \\ \hline
  V & $0\sim255$ & $90\sim240$ \\
  \hline

\end{tabular}
\label{table:haze_detection}
\end{table}

\section{ACCELERATION
OF PROPOSED VIDEO ENHANCEMENT PROCESSING ALGORITHM} \label{sec:opt}

The algorithm described in Section \ref{sec:enh} is a frame based
approach. Through experimental results, we found that the calculation
of $t(x)$ occupies about 60\% of the total computation time.
For real-time and low complexity processing of video
inputs, it is not desirable to apply the algorithm of Section
\ref{sec:enh} on a frame by frame basis, which not only has high computational
complexity, but also makes the output results much more
sensitive to temporal and spatial noise, and destroys the
temporal and spatial consistency of the processed outputs,
thereby lower the overall perceptual quality.

To solve these problems, we notice that the $t(x)$
and other model parameters are correlated temporally and
spatially. Therefore, we propose to accelerate the algorithm by
introducing motion estimation.

Motion estimation/compensation (ME/MC) is a key procedure of the
state-of-the-art video compression standards. By matching blocks in
subsequently encoded frames to find the ``best'' match of a block to
be encoded and a block of the same size that has already been
encoded and then decoded (referred to as the ``reference''), video
compression algorithms use the reference as a prediction of the
block to be encoded and encodes only the difference (termed the
``residual'') between the reference and the block to be encoded,
thereby reducing the rate that is required to encode the current
block to a fidelity level. The process of finding the best match
between a block to be encoded and a block in a reference frame is
called ``motion estimation", and the ``best'' match is usually
determined by jointly considering the rate and distortion costs of
the match. If a ``best'' match block is found, the current block
will be encoded in inter mode and only the residual will be encoded.
Otherwise, the current block will be encoded in intra mode. The most
commonly used metric for distortion in motion estimation is the Sum
of Absolute Differences (SAD).

\begin{figure}[h]
\centering
\includegraphics[width=0.48\textwidth]{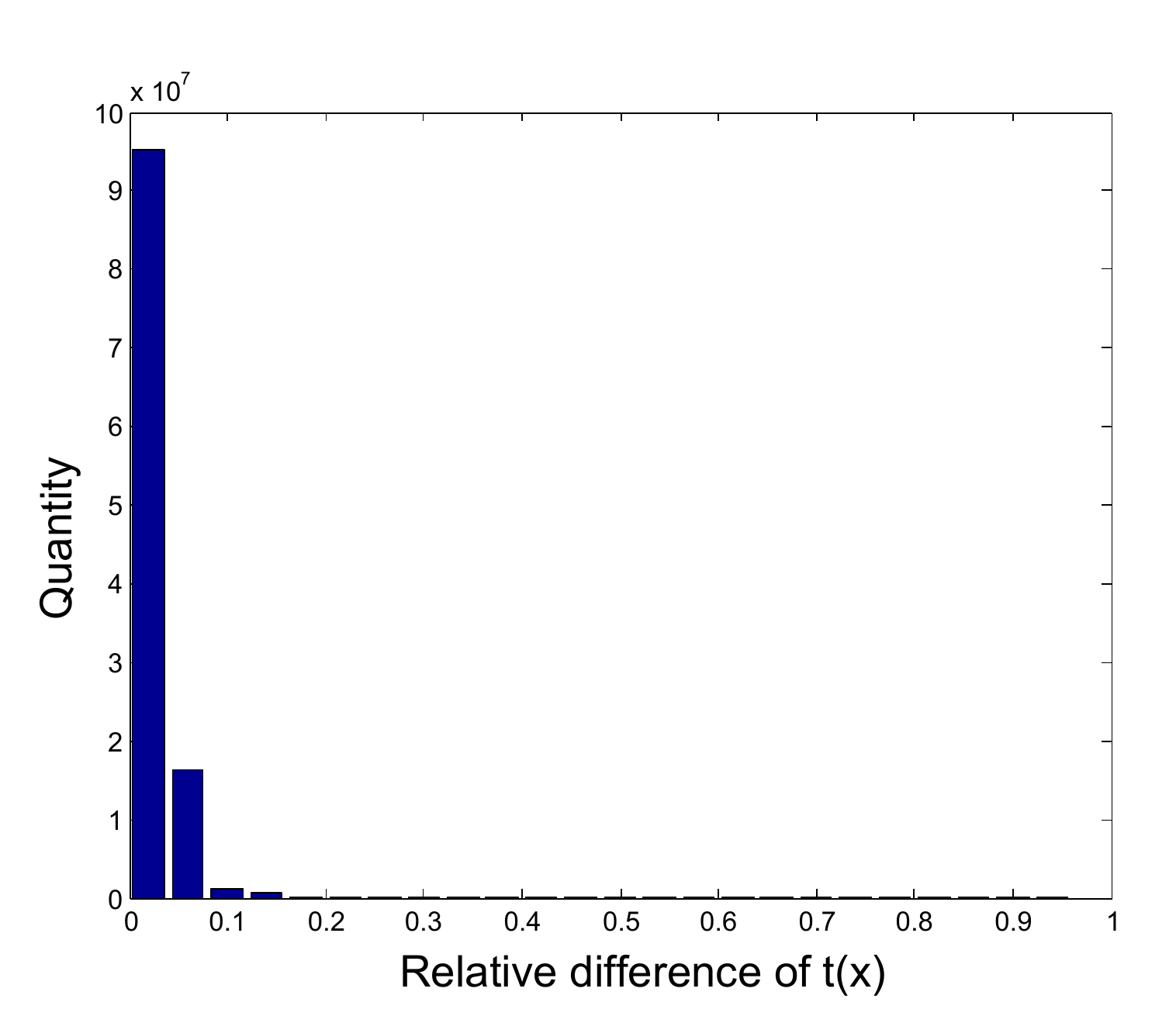}
   \caption{Differences of $t(x)$ values between the predicted block's pixels' and its reference block's pixels'.}
   \label {fig:tx_difference}
\end{figure}

\begin{figure}[h]
\centering
\includegraphics[width=0.25\textwidth]{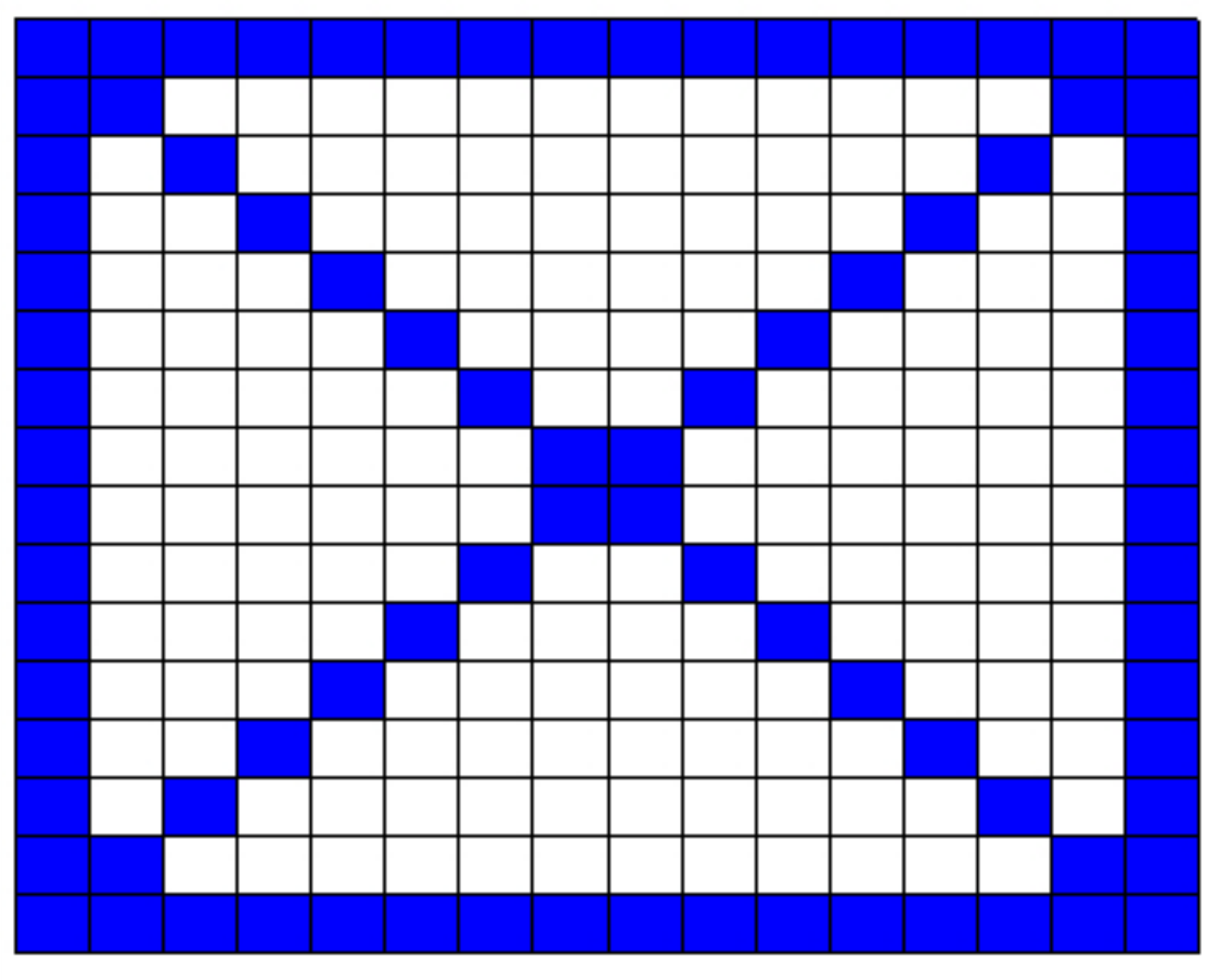}
   \caption{Subsampling pattern of proposed fast SAD algorithm.}
   \label {fig:SAD_pattern}
\end{figure}

To verify the feasibility of using temporal block matching and ME to
expedite $t(x)$ calculation, we calculated the differences of $t(x)$
values for pixels in the predicted and reference blocks. The
statistics in Fig. \ref{fig:tx_difference} shows that the
differences are less than 10\% in almost all cases. Therefore, we
could utilize ME/MC to accelerate the computationally intensive
calculation of $t(x)$ and only needed to calculate $t(x)$ of a few selective
frames. For the non-critical frames, we used the corresponding $t(x)$
values of the reference pixels. To reduce the complexity of the
motion estimation process, we used mature fast motion estimation
algorithms e.g. \emph{Enhanced Prediction Zonal Search (EPZS)}
\cite{EPZS}. When calculating the SAD, similar to  \cite{SAD1} and
\cite{SAD2}, we only utilized a subset of the pixels in the current
and reference blocks using the pattern shown in Fig.
\ref{fig:SAD_pattern}. With this pattern, our calculation ``touched''
 a total of 60 pixels in a $16 \times 16$ block, or roughly 25\%. These pixels
were located on either the diagonal or the edges, resulting in
about 75\%  reduction in SAD calculation when implemented in
software on a general purpose processor.

\begin{figure}[h]
\centering
\includegraphics[width=0.48\textwidth]{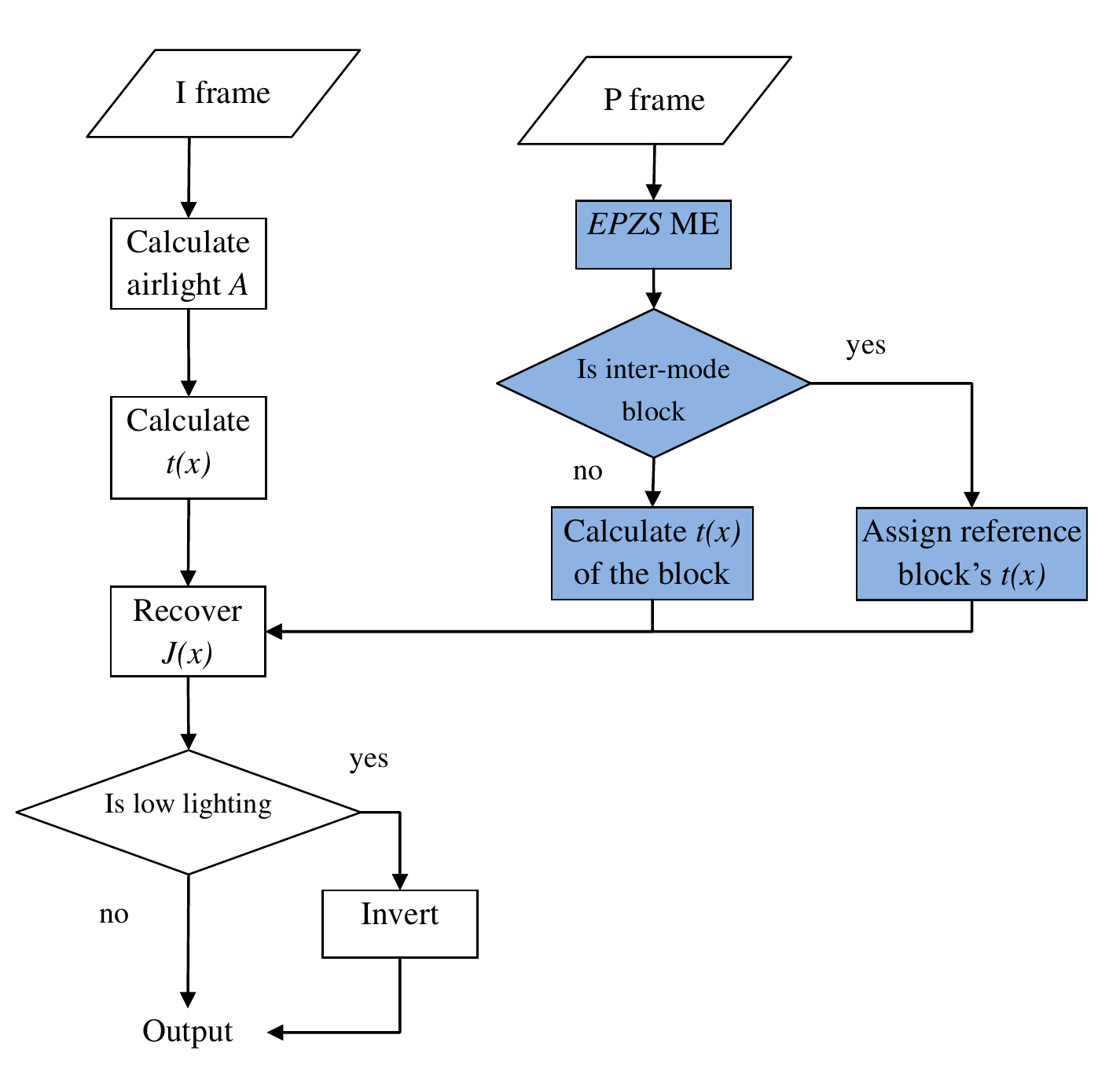}
   \caption{Flow diagram of the core enhancement algorithm with ME acceleration.}
   \label {fig:flow_me}
\end{figure}

In our implementation, when the proposed algorithm is deployed prior
to video compression or after video decompression, we  first divide
the input frames into GOPs. The GOPs could either contain a fix
number of frames, or decided based on a max GOP size (in frames) and
scene changing. Each GOP starts with an Intra coded frame (I frame),
for which all $t(x)$ values are calculated. ME is performed for the
remaining frames (P frames) of the GOP, similar to conventional
video encoding. To this end, each P frame is divided into
non-overlapping $16\times 16$ blocks, for which a motion search
using the SAD is conducted. A threshold $T$ is defined for the SAD
of blocks: if the SAD is below the threshold which means a
``best'' match block is found, the calculation of $t(x)$ for the
entire MB is skipped. Otherwise, $t(x)$ still needs to be
calculated. In both cases, the values for the current frame are
stored for possible use for the next frame. The flow diagram is
shown in Fig. \ref{fig:flow_me}. We call this acceleration algorithm
as \emph{ME acceleration enhancement algorithm}.

\begin{figure*}
\centering
\includegraphics[width=0.9\textwidth,height=0.38\textwidth]{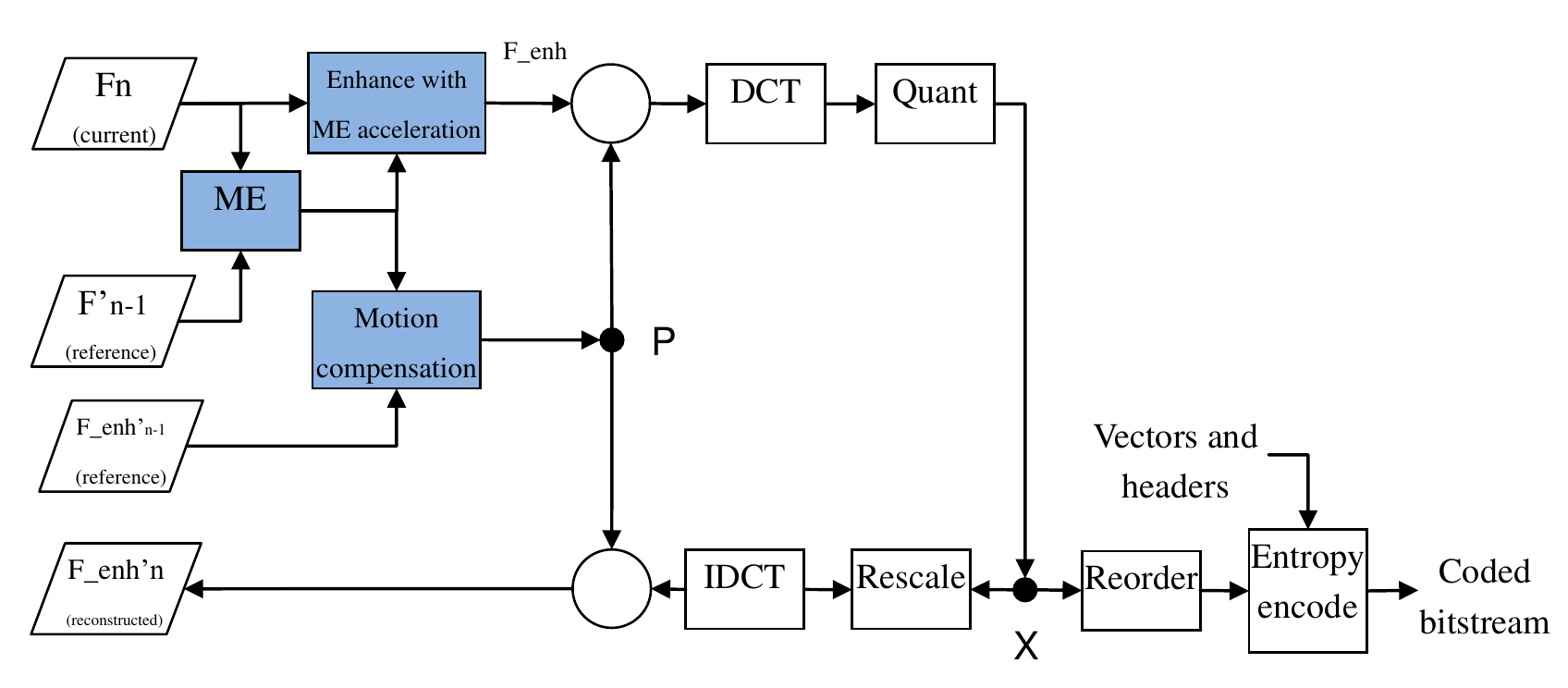}
   \caption{Flow diagram of the integration of encoder and ME acceleration enhancement algorithm.}
   \label {fig:flow_encoder}
\end{figure*}

\begin{figure*}
\centering
\includegraphics[width=0.9\textwidth,height=0.28\textwidth]{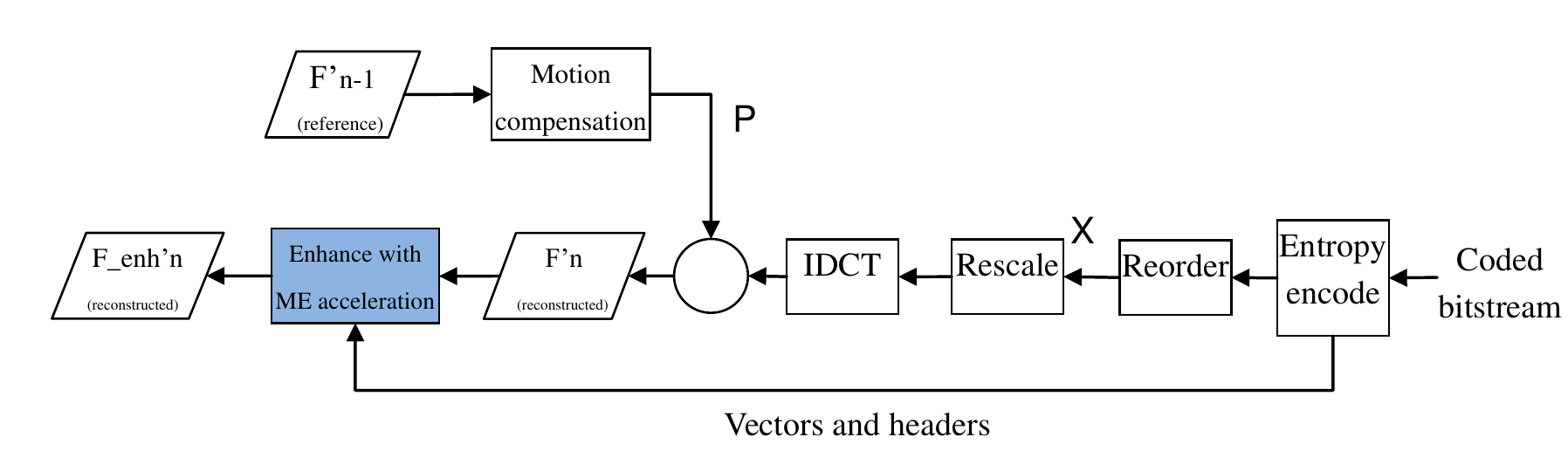}
   \caption{Flow diagram of the integration of decoder and ME acceleration enhancement algorithm.}
   \label {fig:flow_decoder}
\end{figure*}

In addition to operating as a stand-along module with uncompressed
pixel information as both the input and output, the ME accelerated
enhancement algorithm could also be integrated into a video encoder
or a video decoder. When the algorithm is integrated with a video
encoder, the encoder and the enhancement can share the ME module.
When integrated with the decoder, the system has the potential of
using the motion information contained in the input video bitstream
directly, and thereby by-passing the entire ME process. Such
integration will usually lead to a RD loss. The reason for this loss
is first and foremost that the ME module in the encoder with which
the enhancement module is integrated or the encoder with which the
bitstreams that a decoder with enhancement decodes may not be
optimized for finding the best matches in $t(x)$ values. For
example, when the enhancement module is integrated with an decoder,
it may have to decode an input bitstream encoded by a low complexity
encoder using a really small ME range. The traditional SAD or
SAD-plus-rate metrics for ME are also not optimal for $t(x)$ match
search. However, through extensive experiments with widely used
encoders and decoders, we found that such quality loss were usually
small, and well-justified by the savings in computational cost. The
flow diagrams of integrating the ME acceleration enhancement
algorithm into encoder and decoder are shown in Fig.
\ref{fig:flow_encoder} and Fig. \ref{fig:flow_decoder}. Some of the
comparisons can be found in Section \ref{sec:results}.

\section{Experimental Results} \label{sec:results}

To evaluate the proposed algorithm, a series of experiments were
conducted with a Windows PC (Intel Core 2 Duo
processor running at 2.0 GHz with 3G of RAM) and an iPhone 4. The resolution of
testing videos in our experiments was $640\times480$.

\begin{figure}[h]
\centering
\includegraphics[width=0.35\textwidth,height=0.25\textwidth]{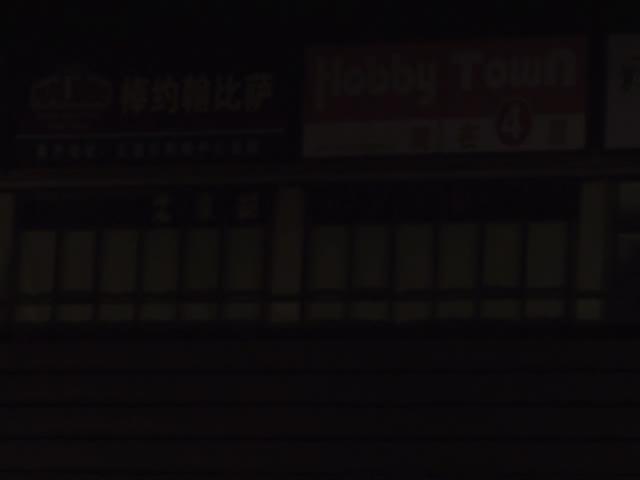}
\includegraphics[width=0.35\textwidth,height=0.01\textwidth]{blank.jpg}
\includegraphics[width=0.35\textwidth,height=0.25\textwidth]{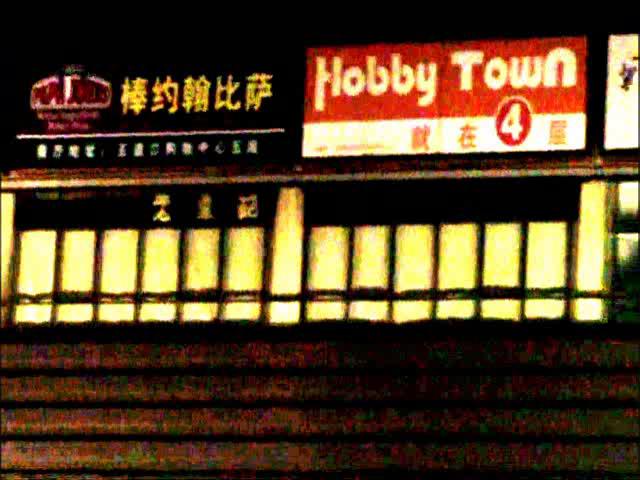}
   \caption{Examples of low lighting video enhancement algorithm: Original input (Top), and the enhancement result (Bottom).}
   \label {fig:result_low_lighting}
\end{figure}

\begin{figure}[h]
\centering
\includegraphics[width=0.35\textwidth,height=0.25\textwidth]{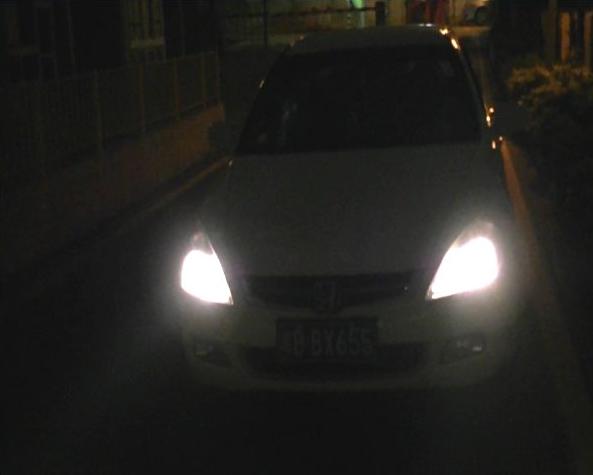}
\includegraphics[width=0.35\textwidth,height=0.01\textwidth]{blank.jpg}
\includegraphics[width=0.35\textwidth,height=0.25\textwidth]{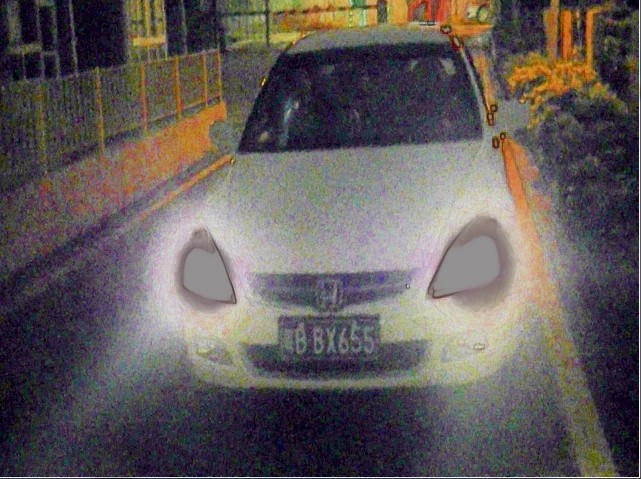}
   \caption{Examples of high dynamic range video enhancement algorithm: Original input (Top), and the enhancement result (Bottom).}
   \label {fig:result_high_lighting}
\end{figure}

\begin{figure}[h]
\centering
\includegraphics[width=0.35\textwidth,height=0.25\textwidth]{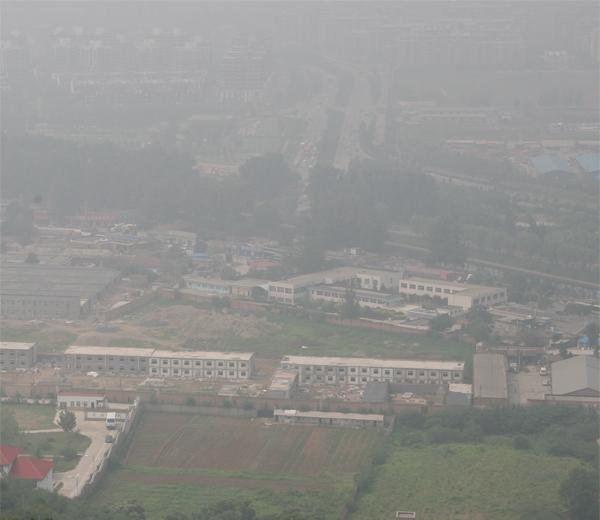}
\includegraphics[width=0.35\textwidth,height=0.01\textwidth]{blank.jpg}
\includegraphics[width=0.35\textwidth,height=0.25\textwidth]{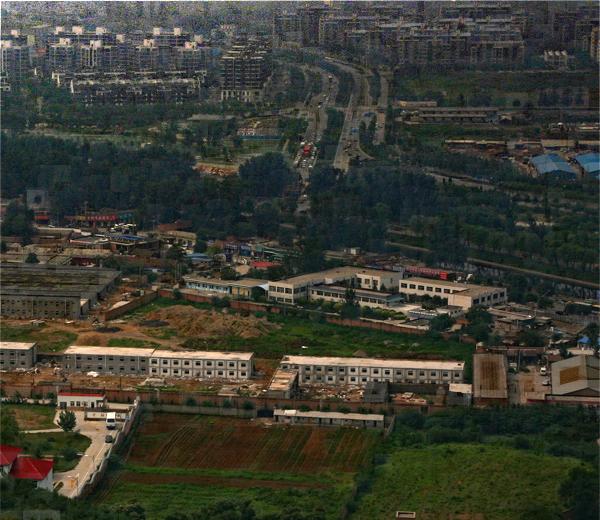}
   \caption{Examples of haze removal algorithm: Original input (Top), and the enhancement result (Bottom).}
   \label {fig:result_haze}
\end{figure}

\begin{figure}[h]
\centering
\includegraphics[width=0.35\textwidth,height=0.25\textwidth]{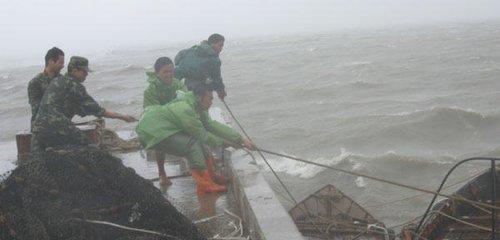}
\includegraphics[width=0.35\textwidth,height=0.01\textwidth]{blank.jpg}
\includegraphics[width=0.35\textwidth,height=0.25\textwidth]{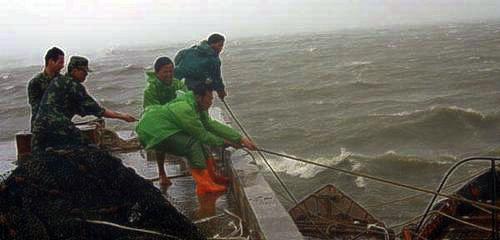}
   \caption{Examples of rainy video enhancement using haze removal algorithm: Original input (Top), and the enhancement result (Bottom).}
   \label {fig:result_rain}
\end{figure}

\begin{figure}[h]
\centering
\includegraphics[width=0.35\textwidth,height=0.25\textwidth]{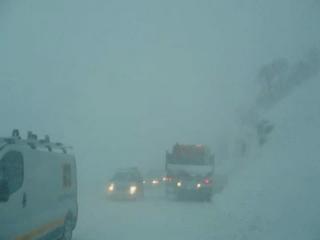}
\includegraphics[width=0.35\textwidth,height=0.01\textwidth]{blank.jpg}
\includegraphics[width=0.35\textwidth,height=0.25\textwidth]{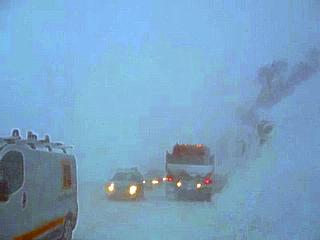}
   \caption{Examples of snowy video enhancement using haze removal algorithm: Original input (Top), and the enhancement result (Bottom).}
   \label {fig:result_snow}
\end{figure}

Examples of the enhancement outputs for low lighting, high dynamic
range and hazy videos are shown in Fig.
\ref{fig:result_low_lighting}, Fig. \ref{fig:result_high_lighting}
and Fig. \ref{fig:result_haze} respectively. As we can see from
these figures, the improvements in visibility are obvious. In Fig. \ref{fig:result_low_lighting}, the
yellow light from the windows and signs such as
``Hobby Town" and other Chinese characters were recovered in correct
color. In
Fig. \ref{fig:result_high_lighting}, the headlight of the car in the original input
made letters on the license plate very difficult to read.
After enhancement with our algorithm,
the license plate became much more intelligible. The algorithm also
worked well for video captured in hazy, rainy and snowy weathers
as shown in Fig.
\ref{fig:result_haze}, Fig.
\ref{fig:result_rain} and Fig. \ref{fig:result_snow}.

\begin{figure}[h]
\centering
\includegraphics[width=0.48\textwidth,height=0.35\textwidth]{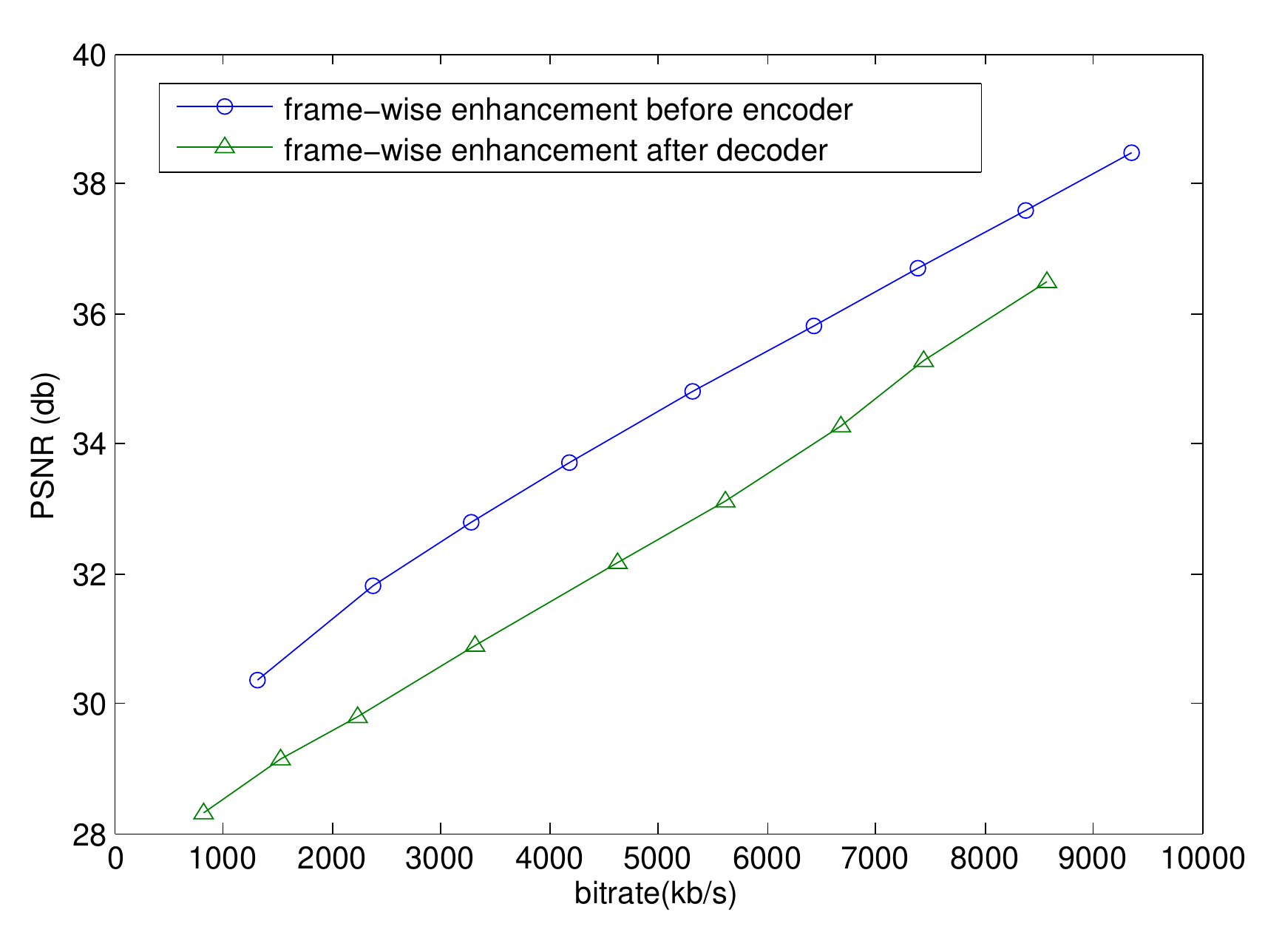}
   \caption{RD performance of frame-wise enhancement in encoder and decoder.}
   \label {fig:traffic_frame_wise}
\end{figure}

\begin{figure}[h]
\centering
\includegraphics[width=0.48\textwidth,height=0.35\textwidth]{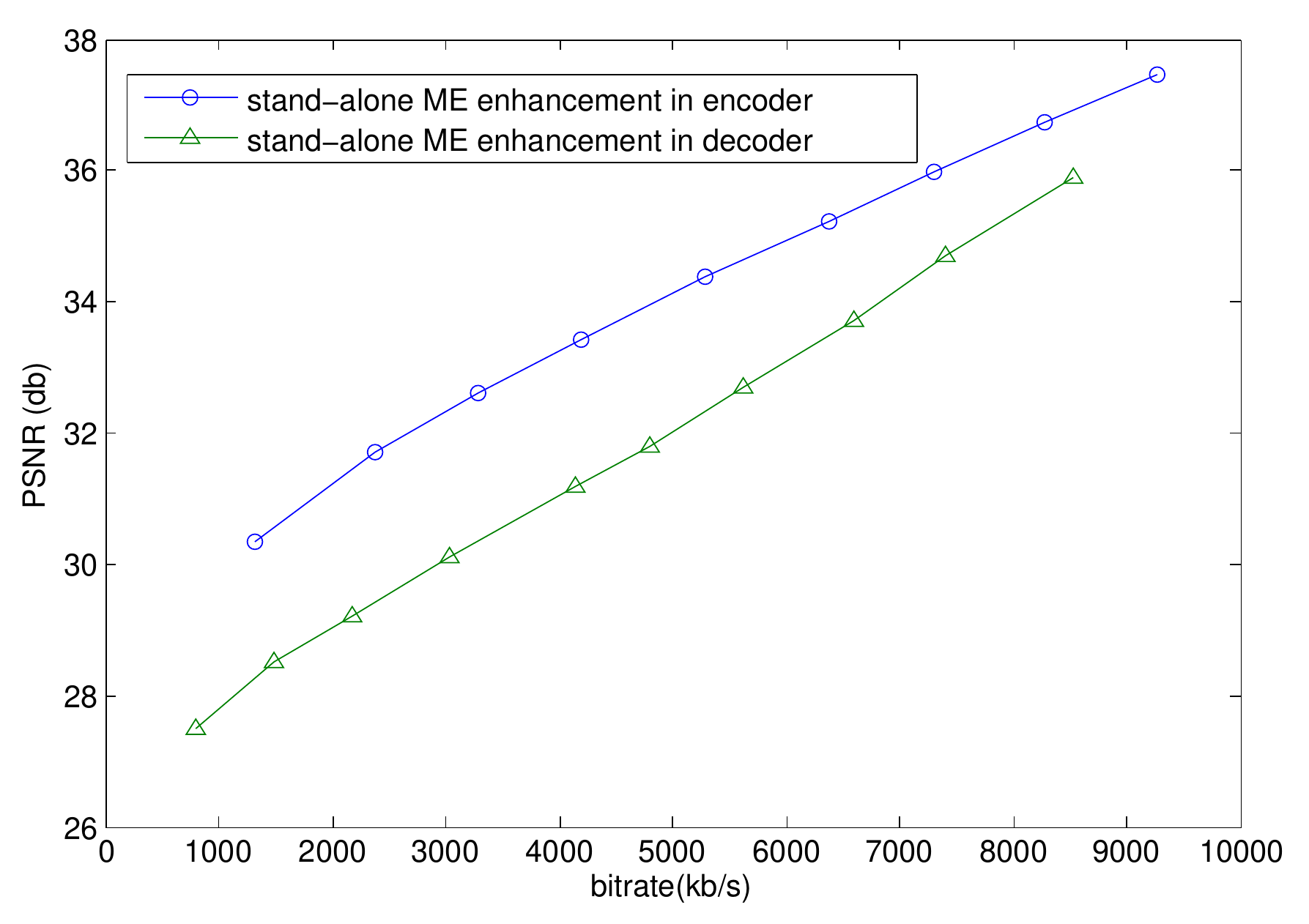}
   \caption{RD performance of separate ME acceleration enhancement in encoder and decoder.}
   \label {fig:traffic_stand_alone}
\end{figure}

\begin{figure}[h]
\centering
\includegraphics[width=0.48\textwidth,height=0.35\textwidth]{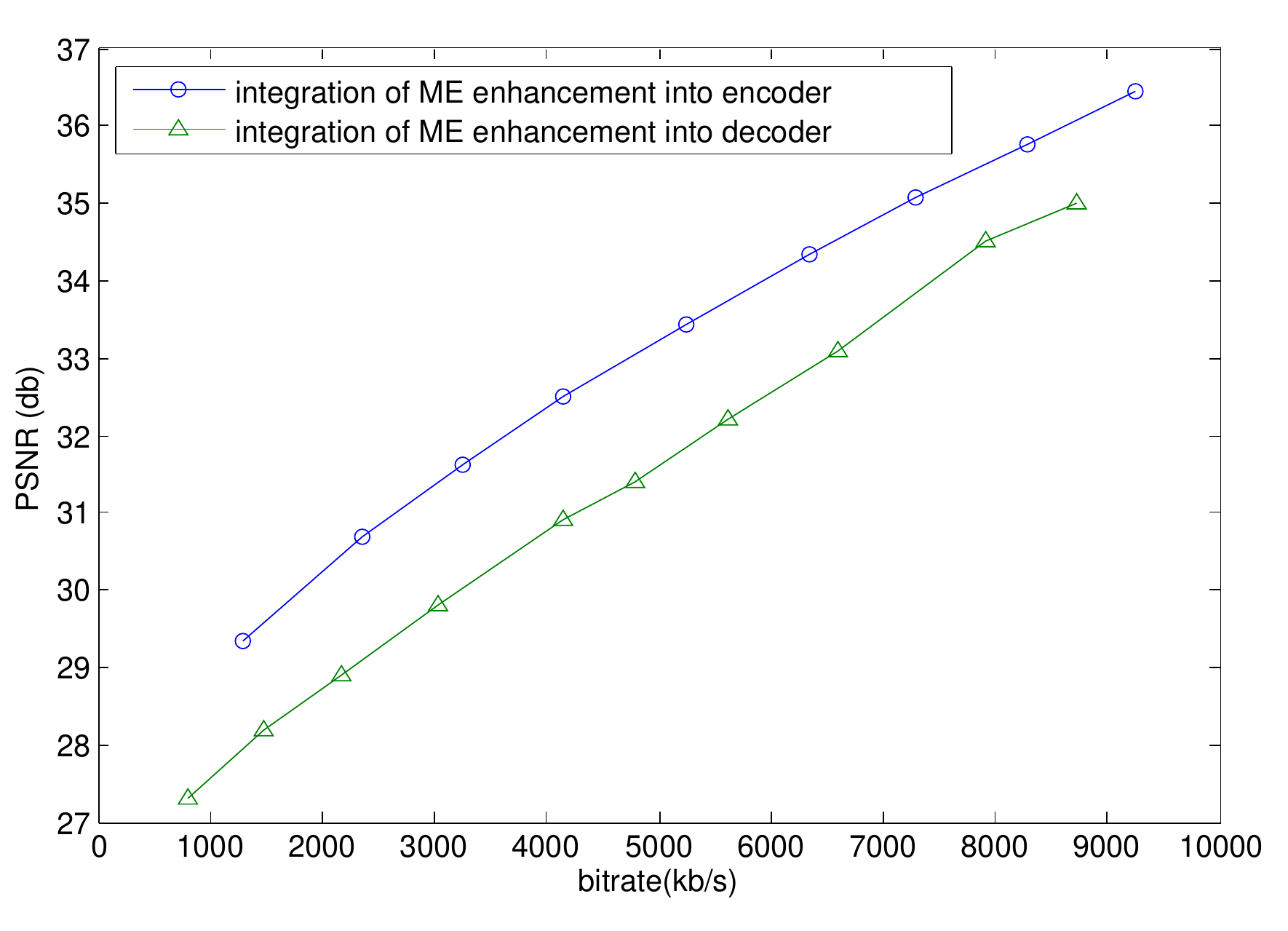}
   \caption{RD performance of integration of ME acceleration enhancement into encoder and decoder.}
   \label {fig:traffic_integration}
\end{figure}

\begin{figure}
\centering
\includegraphics[width=0.15\textwidth,height=0.13\textwidth]{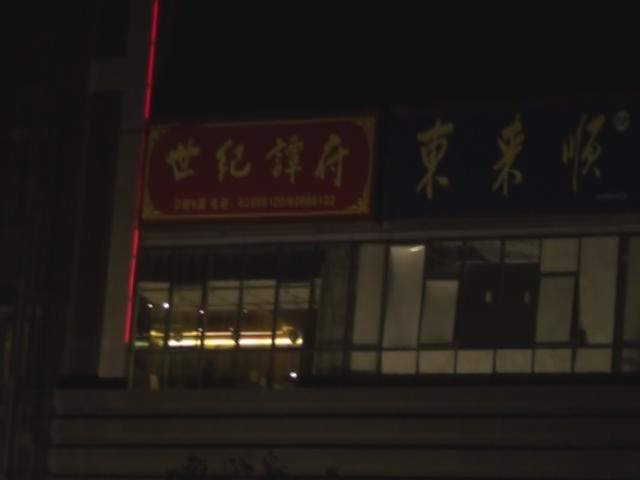}
\includegraphics[width=0.15\textwidth,height=0.13\textwidth]{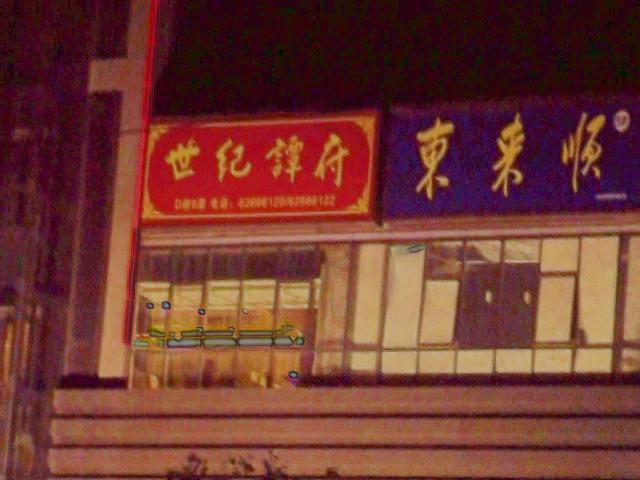}
\includegraphics[width=0.15\textwidth,height=0.13\textwidth]{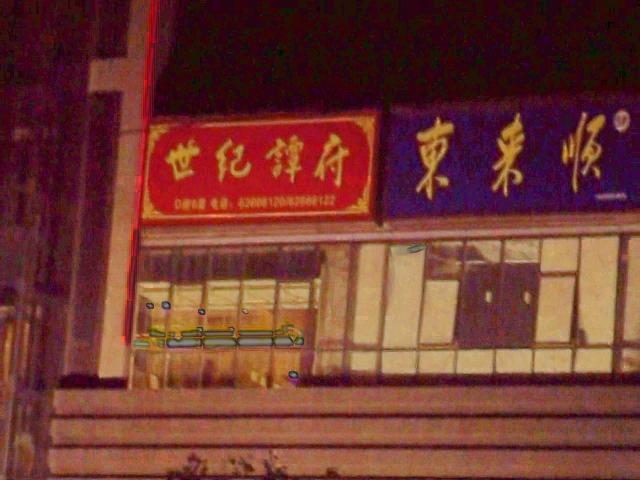}
\includegraphics[width=0.35\textwidth,height=0.01\textwidth]{blank.jpg}
\includegraphics[width=0.15\textwidth,height=0.13\textwidth]{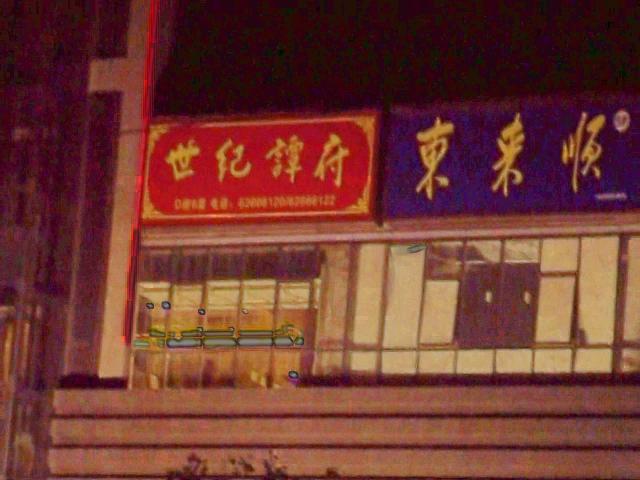}
\includegraphics[width=0.15\textwidth,height=0.13\textwidth]{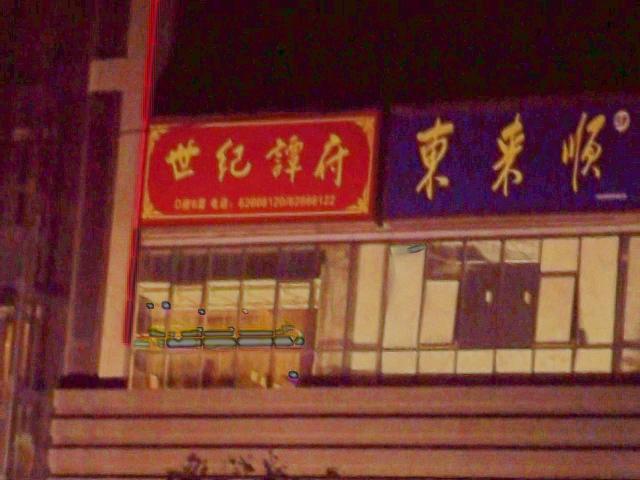}
   \caption{Examples of comparisons among the frame-wise algorithm and the three proposed ME acceleration methods: Original input (Top left), output of frame-wise algorithm (Top middle), output of separate ME acceleration algorithm (Top right), output of integration of ME acceleration algorithm into encoder (Bottom left) and decoder (Bottom right).}
   \label {fig:compare}
\end{figure}

In addition, the proposed ME-based acceleration greatly reduces the
complexity of the algorithm with little information lost. As
mentioned above, there are three possible ways of incorporating ME
into the enhancement algorithm, i.e. through a separate ME module in
the enhancement system, as well as utilizing the ME module and
information available in a video encoder or decoder. Some example
outputs of the frame-wise enhancement algorithm and these three ways
of incorporating ME are shown in Fig. \ref{fig:compare}, with
virtually no visual difference. We also calculated the average RD
curves of ten randomly selected experimental videos using the three
acceleration methods. The reference was enhancement using the
proposed frame-wise enhancement algorithm in YUV region. The RD
curves of performing the frame-wise enhancement algorithm before
encoding or after decoding are shown in Fig.
\ref{fig:traffic_frame_wise}, while the results for acceleration
using a separate ME module are given in Fig.
\ref{fig:traffic_stand_alone}, and integrating the ME acceleration
into the codec are shown in Fig. \ref{fig:traffic_integration}. As
the RD curves in our experiments reflect the aggregated outcome of
both coding and enhancement, and because enhancement was not
optimized for PSNR based distortion, the shape of our RD curve looks
different from RD curves for video compression systems, even though
distortion as measured in PSNR is still a monotonic function of the
rate. First, from the three figures, we find that in general,
performing enhancement before encoding has better overall RD
performance. Although enhancing after decoding means we can transmit
un-enhanced video clips, which usually having lower contrast, less
detail and are easier to compress, the reconstructed quality after
decoder/enhancement is heavily affected by the loss of quality
during the encoding, leading to an overall RD performance loss of 2
dB for the cases in the experiments. In addition, in Fig.
\ref{fig:traffic_frame_wise}, the RD loss of frame-wise enhancement
was due to encoding and decoding. In Fig.
\ref{fig:traffic_stand_alone}, the RD loss resulted from ME
acceleration and encoding/decoding. In Fig.
\ref{fig:traffic_stand_alone}, the RD loss resulted from integration
of ME acceleration algorithm into encoder and decoder. Overall
however, the RD loss introduced by ME acceleration and integration
was small in PSNR terms, and not visible subjectively.

\begin{table*}
\caption {Processing speeds of proposed algorithms over PC and
iPhone4} \centering
\begin{tabular}{|c|c|c|c|}
  \hline
    &PC/ms per frame&iPhone4/ms per frame&Time saved\\ \hline
  Frame-wise enhancement algorithm&40.1& 500.3&N/A\\ \hline
  Separate ME acceleration enhancement algorithm&29.3& 369&27.5\%\\ \hline
  Integration of ME acceleration enhancement algorithm into encoder&9.2& 107.9&77.3\%\\ \hline
  Integration of ME acceleration enhancement algorithm into decoder&24.8& 302.4&40.0\%\\ \hline
\end{tabular}
\label{table:result_speed}
\end{table*}

We also measured the computational complexity of frame-wise
enhancement, acceleration with a separate ME module and integration
into an encoder or a decoder. The computational cost was measured in
terms of average time spent on enhancement per frame. For the cases
when the enhancement was integrated into the codec, we did not count
the actual encoding or decoding time, so as to measure only the
enhancement itself.  As shown in the Table \ref{table:result_speed},
using a separate ME module saved about 27.5\% time on average
compared with the frame-wise algorithm. On the other hand,
integrating with the decoder saved 40\% time compared with the frame
wise algorithm, while integrating with the encoder saved about
77.3\%.

\section{Conclusions} \label{sec:conclusions}
In the paper, we propose a novel fast and efficient integrated
algorithm for real-time enhancement of videos acquired under
challenging lighting conditions including low lighting, bad weather (hazy,
rainy, snowy) and
high dynamic range conditions. We show that visually and statistically, hazy
video and video captured in various challenging lighting conditions
are very similar, and therefore a single core enhancement algorithm
can be utilized in all cases, along with a proper pre-processing and
an automatic impairment source detection module. We also describe
a number of ways of reducing the computational complexity of
the system while maintaining good visual quality, and the tradeoffs involved
when the proposed system is integrated into different modules of the
video acquisition, coding, transmission and consumption chain.

Areas of further improvements include better pre-processing filters
targeting specific sources of impairments, improved core enhancement
algorithm, and better acceleration techniques. Also of great importance
is a system that can process inputs with compounded impairments (e.g.
video of foggy nights, with both haze and low lighting).
\section{Acknowledgments}
The authors wish to thank the SRT students Yao Lu, Wei Meng, Yuanjie
Liu, Huan Jing and Xuefeng Hu at the Media Lab of the Department of
Computer Science and Technology of Tsinghua University for the help
they provided to this paper.

\ifCLASSOPTIONcaptionsoff
  \newpage
\fi


\begin{thebibliography}{1}

\bibitem {FIR1} M. Blanco, H. M. Jonathan, and T. A. Dingus. ``Evaluating New Technologies to Enhance Night Vision by Looking at Detection and Recognition Distances of Non-Motorists and Objects,''
in \emph{Proc. Human Factors and Ergonomics Society}, Minneapolis,
MN, Jan. 2001, vol. 5, pp. 1612-1616.


\bibitem {FIR2} O. Tsimhoni, J. B\"{a}rgman, T.
Minoda, and M. J. Flannagan. ``Pedestrian Detection with Near and
Far Infrared Night Vision Enhancement,'' Tech. rep., The University
of Michigan, 2004.


\bibitem{NIR1} L. Tao, H. Ngo, M. Zhang, A. Livingston, and V. Asari. ``A Multi-sensor Image Fusion and Enhancement System for Assisting Drivers in Poor Lighting Conditions,''
in \emph{Proc. IEEE Conf. Applied Imagery and Pattern Recognition
Workshop}, Washington, DC, Dec. 2005, pp. 106-113.





\bibitem{NIR2} H. Ngo, L. Tao, M. Zhang, A. Livingston, and V.
Asari. ``A Visibility Improvement System for Low Vision Drivers by
Nonlinear Enhancement of Fused Visible and Infrared Video,'' in
\emph{Proc. IEEE Conf. Computer Vision and Pattern Recognition}, San
Diego, CA, Jun. 2005, pp.25.







\bibitem{ICCV07} H. Malm, M. Oskarsson, E. Warrant, P. Clarberg, J. Hasselgren, and C. Lejdfors. ``Adaptive Enhancement and Noise Reduction in Very Low Light-Level Video,''
 in
\emph{Proc. IEEE Int. Conf. Computer Vision}, Rio de Janeiro,
Brazil, Oct. 2007, pp. 1-8.



\bibitem{Siggraph05} E. P. Bennett, L. McMillan. ``Video Enhancement Using Per-pixel Virtual Exposures,'' in \emph{Proc. SIGGRAPH '05}, Los Angeles, CA,
Jul. 2005, pp. 845-852.


\bibitem{1924} Koschmieder. ``Theorie der horizontalen sichtweite,''
in \emph{Beitr. Phys. Freien Atm.}, vol. 12, pp. 171-181, 1924.



\bibitem{R.Lim} R. Lim, T. Bretschneider. ``Autonomous Monitoring of Fire-related Haze from Space,''
 in
\emph{Conf. Imaging Science, Systems and Technology}, Las Vegas,
Nevada, Jun. 2004, pp. 101-105.

\bibitem{Song} C. Song, C. E. Woodcock, K. C. Seto, M. P. Lenney, and S. A.
Macomber. ``Classification and Change Detection Using Landsat TM
Data: When and How to Correct Atmospheric Effects?'' in \emph{Int.
Symposium Remote Sensing of Environment.}, vol. 75, no. 2, pp.
230-244, Feb. 2001.

\bibitem{Yong} Du Y., Guindong B., and Cihlar J.. ``Haze Detection and
Removal in High Resolution Satellite Image with Wavelet Analysis,''
in \emph{IEEE Trans. Geoscience and Remote Sensing}, vol. 40, no. 1,
pp. 210-217, Jan. 2002.

\bibitem{Chi} R.A. Fisher, and F. Yates. ``Statistical Tables for Biological, Agricultural and Medical Research,'' in 6th Ed. Oliver and Boyd, Ltd., Edinburgh and London,
1963, pp. 10-30.

\bibitem{He} K. He, J. Sun, and X. Tang. ``Single Image Haze
Removal Using Dark Channel Prior,'' in \emph{Proc. IEEE Conf.
Computer Vision and Pattern Recognition}, Miami, FL, Jun. 2009, pp.
1956-1963.

\bibitem{Fattal} R. Fattal. ``Single Image Dehazing,'' in \emph{Proc. SIGGRAPH '08}, Los Angeles, CA,
Aug. 2008, pp. 1-9.


\bibitem{Tan} R. Tan. ``Visibility in Bad Weather from A Single
Image,'' in \emph{Proc. IEEE Conf. Computer Vision and Pattern
Recognition}, Anchorage, Alaska, Jun. 2008, pp. 1-8.

\bibitem{Na} S. G. Narasimhan, and S. K. Nayar. ``Chromatic Framework for
Vision in Bad Weather,'' in \emph{Proc. IEEE Conf. Computer Vision
and Pattern Recognition}, Hilton Head, SC, Jun. 2000, vol. 1, pp.
1598-1605.


\bibitem{EPZS} A. M. Tourapis. ``Enhanced Predictive Zonal Search for Single and Multiple Frame Motion Estimation,''
in \emph{Proc. Visual Communications and Image Processing}, San
Jose, CA, Jan. 2002, pp. 1069-1079.


\bibitem{SAD1} T. Koga, K. Iinuma, A. Hirano, Y. Iijima, and T. Ishiguro.``Motion
Compensated Interframe Coding for Video Conferencing,'' in
\emph{Proc. Nut. Telecommun. Conf.}, New Orleans, LA, Nov. 1981, pp.
G5.3.1-G5.3.5.

\bibitem{SAD2} B. Girod, and K. W. Stuhlm\"{u}ller. ``A
Content-Dependent Fast DCT for Low Bit-Rate Video Coding,'' in
\emph{Proc. IEEE Int. Conf. Image Process.}, Chicago, Illinois, Oct.
1998, vol. 3, pp. 80-83.








\bibitem{Xuan} X. Dong, Y. Pang, and J. Wen. ``Fast Efficient Algorithm for Enhancement of Low Lighting Video,'' \emph{SIGGRAPH '10 Poster}, Los Angeles, CA, Jul.
2010.

\bibitem{Xuan2} X. Dong, Y. Pang, J. Wen, G. Wang, W. Li, Y. Gao, and S. Yang.
``A Fast Efficient Algorithm for Enhancement of Low Lighting
Video,'' in \emph{Int. Conf. Digital Media and Digital Content
Management}, Chongqing, China, Dec. 2010.

\end{thebibliography}
\end{document}